\documentclass[11pt,a4paper]{article}

\usepackage{amsmath}
\usepackage{amssymb}
\usepackage{multirow}
\usepackage{graphicx}
\usepackage{mathptmx}
\usepackage{natbib}
\usepackage{floatrow}
\usepackage[caption=false]{subfig}
\usepackage[usenames,dvipsnames]{color}
\usepackage{verbatim}
\usepackage[margin=1.15in]{geometry}

\begin{document}

\title{Complete eddy self-similarity in turbulent pipe flow}
\author{L. H. O. Hellstr\"om, T. Van Buren$^{1}$, J. C. Vaccaro$^{2}$ and A. J. Smits$^{3}$\\
$^1$Mechanical Engineering, University of Delaware, U.S.A\\
$^2$Mechanical and Aerospace Engineering, Princeton University, U.S.A.\\
$^3$Department of Engineering, Hofstra University, U.S.A.\\}

\date{\today}
\floatsetup[figure]{style=plain,subcapbesideposition=top}

\maketitle

%------------------------------------------------------
% Abstract --------------------------------------------
%------------------------------------------------------
\begin{abstract}

For wall-bounded turbulent flows, Townsend's attached eddy hypothesis proposes that the logarithmic layer is populated by a set of energetic and geometrically self-similar eddies. These eddies scale with a single length scale, their distance to the wall, while their velocity scale remains constant across their size range. To investigate the existence of such structures in fully developed turbulent pipe flow,  stereoscopic particle image velocimetry measurements were performed in two parallel cross-sectional planes, spaced apart by a varying distance from 0 to 9.97$R$, for $Re_\tau = 1310$, 2430 and 3810.  

The instantaneous turbulence structures are sorted by width using an azimuthal Fourier decomposition, allowing us to create a set of average eddy velocity profiles by performing an azimuthal alignment process. The resulting eddy profiles exhibit geometric self-similar behavior in the $(r,\theta)$-plane for eddies with spanwise length scales ($\lambda_\theta/R$) spanning from 1.03 to 0.175. The streamwise similarity is further investigated using two-point correlations between the two planes, where the structures exhibit a self-similar behaviour with length scales ($\lambda_\theta/R$) ranging from approximately $0.88$ to $0.203$. The candidate structures thereby establish full three-dimensional geometrically self-similarity for structures with a volumetric ratio of $1:80$. Beside the geometric similarity, the velocity magnitude also exhibit self-similarity within these ranges. 
However, the velocity scale depends on eddy size, and follow the trends based on the scaling arguments proposed by \cite{Perry1986}.

\end{abstract}

%------------------------------------------------------
% Introduction ----------------------------------------
%------------------------------------------------------

\section{Introduction}

In the search of a physics-based model of wall-bounded turbulence, \cite{Townsend1976} introduced the concept that the constant stress region of a wall-bounded flow is populated by geometrically self-similar, inertial eddies. The velocity field induced by each of these eddies is taken to extend to the wall and, in this sense they are attached to the wall.  That is, each individual eddy is aware of the wall, and scales solely by its distance to it.  They are not physically connected to the wall because in close proximity to the wall viscosity must become important. This  ``attached eddy hypothesis'' provides a simple model of wall-bounded turbulent flows, although it is at heart an inviscid flow model. 

The model assumes a linear superposition of self-similar, attached eddies that span a wide range of scales. The actual range of scales is defined by the Reynolds number, and it grows as the Reynolds number increases.  Townsend further prescribed that the velocity field of each eddy is scaled by a constant characteristic velocity scale, the friction velocity, and that the probability distribution of each size is inversely proportional to the distance to the wall.  With these prescriptions, the model produces a constant shear stress region corresponding to the logarithmic region, and it  yields the logarithmic mean flow profile, along with the variances and higher order moments. 

\cite{Perry1986} argued that there is an overlap region in wavenumber space ($k$) between eddies that scale with their wall-normal distance ($y$) and eddies that scale with the outer length scale ($R$). Under these circumstances the power spectra density function would exhibit a $k^{-1}$ slope in the log-log spectrum, and a plateau in the pre-multiplied spectrum. When the power spectrum is integrated (for high Reynolds numbers) the streamwise and spanwise Reynolds stress components exhibit a logarithmic behaviour with respect to $y$, while the wall-normal component is reduced to a constant. These scaling arguments at very high Reynolds numbers were most recently investigated by \cite{vallikivi2015} whom found no clear evidence for a $k^{-1}$ slope or a plateau, while the streamwise Reynolds stress exhibited a clear logarithmic behaviour with respect to $y$. However, if the proposed scaling for the overlap region in wavenumber space is assumed to be true, the eddy velocity field would scale as $ y u_\tau$.
A more detailed reading regarding Townsend's attached eddy hypothesis is given by \cite{Marusic2019}.

Much effort has gone into finding support for the attached eddy hypothesis, primarily by verifying the predictions for the Reynolds stresses. For instance, experiments by \cite{Hultmark2012} and \cite{Marusic2013} give strong support for the predicted logarithmic variation of streamwise component at high Reynolds number, while direct numerical computations (DNS) by \cite{Jimenez2008} and \cite{Lee_Moser2015} show that the predictions hold for the wall-parallel and wall-normal components at considerably lower Reynolds numbers. \cite{Hwang2018} further showed that the probabilities of the eddy populations are inversely proportional to their length scale.

In contrast, the evidence for the presence of self-similar structures themselves is fairly limited. In the initial work, \cite{Townsend1976} assumed that the near-wall streaks found by \cite{Kline1967} were self-similar, and modeled the statistical eddy as a double cone vortex. \cite{Perry1982} and \cite{Perry1986} used the observations by \cite{Head1981}, and modeled the self-similar eddies as hairpin vortices. Because the attached eddy model is a linear model, all non-linear interactions between eddies need to be either small, or included within each eddy profile. In this respect, \cite{Adrian2000} showed that hairpin vortices of different sizes tend to align in the streamwise direction, creating the so-called hairpin packets or large-scale motions (LSMs). The LSMs can be thought to contain all non-linear interactions, while the interaction between LSMs is linear. \cite{Marusic2001} and \cite{Woodcock2015} used these findings to generate an aggregated self-similar structure, where all non-linear interactions are contained in the structure itself.

The inertial eddies are expected to be the energetically dominant structures.  One useful approach to identifying and analyzing energetic structures in turbulence is to decompose the flow into basis functions using Proper Orthogonal Decomposition (POD), where each function is ordered based on its energy contribution.  For instance, \cite{Hellstrom2015} studied fully developed turbulent pipe flow at $Re_\tau= u_\tau R/\nu=2460$, where $u_\tau = \sqrt{\tau_w /\rho}$, $\tau_w$ is the wall shear stress, $R$ is the pipe radius, $\rho$ is the fluid density,  and $\nu$ is its kinematic viscosity. They acquired high-speed particle image velocimetry (PIV) data simultaneously in two orthogonal planes, resolving the streamwise $(r,x)$ and cross-stream $(r,\theta)$ planes, where the temporal analysis of the cross-stream plane revealed large structures with a spatio-temporal extent of $1\hbox{-}2R$. These structures appeared to have characteristics very similar to the LSM, which was further supported by combining the analysis of the two planes and creating a conditionally-averaged structure based on the occurrence/intensity of a given cross-stream snapshot POD mode. The resulting structures consisted of a combination of wall-attached and wall-detached large-scale components, which were shown to be associated with the most energetic modes. A pseudo-alignment of these structures was observed, that together created structures with a spatio-temporal extent of about $6R$, which appears to be in accord with the suggestion by \cite{Kim1999} that the very large scale motions are not spatial structures but a temporal manifestation of repeating LSMs.
Furthermore, the method of using a Fourier decomposition in the azimuthal direction has been shown to be exceptionally well suited for finding large coherent structures in pipe flow. For instance, \cite{Hellstrom2015} performed a PIV study using a cross-plane and a streamwise-radial plane simultaneously. They created a conditional structure based on the POD coefficients found in the azimuthally decomposed cross-plane. These structures were identified to be the LSMs. Later \cite{Hellstrom2017} performed a similar structure identification method on DNS data, and found  the LSM velocity footprint as well as the associated pressure field. An important feature of the conditional mode is that, although the flow is no longer forced to be azimuthally periodic, the POD modes are only used to create the signal that conditions the averaging process. Consequently, the structures are still aligned next to one another, showing that the azimuthal structure alignment is an inherit feature of the flow.

\cite{Hellstrom2016a} used similar experimental data as \cite{Hellstrom2015}, but explored the scaling of the POD modes in terms of Townsend's attached eddy hypothesis. As a consequence of the available flow symmetries in pipe flow, they decomposed the azimuthal direction into Fourier modes, where the spanwise wavelength $\lambda_\theta$ defined the width of each structure. The radial direction was decomposed using POD, and sorted by their energy contribution. For two Reynolds numbers, $Re_\tau = 1330$ and $2460$, they found that the first three POD modes (or eddies) exhibited self-similar behavior with respect to their azimuthal wavelengths, with their wall-normal length scales spanning a decade. This single length scale, derived from the azimuthal scaling, provided a complete description of the cross-sectional shape of the self-similar eddies. However, it is not yet clear to what degree the higher order modes are self-similar, nor do we know which higher-order modes may be truncated altogether. In addition, although POD analysis is an excellent tool for identifying and analyzing geometrical self-similarity, it is not as useful for studying the scaling of the complete velocity field.  

Therefore, we now extend this work to address the full three-dimensional similarity of the eddy motions, and their velocity scaling. The work considers a dual cross-plane PIV setup, where the distance between the two planes was varied in 21 steps, ranging from 0 to $9.97R$. The analysis is performed at three Reynolds numbers ($Re_\tau \in \{ 1310,~2430,~3810\}$) and can be broken down into four parts: i) deriving a representative structure; ii) azimuthal/wall-normal similarity; iii) velocity scaling; and iv) azimuthal/streamwise similarity.

%------------------------------------------------------
% Experimental setup ----------------------------------
%------------------------------------------------------
\section{Experimental setup}

\begin{figure}
\vspace{2mm}
 	\begin{center}
		\includegraphics[width=0.6\textwidth]{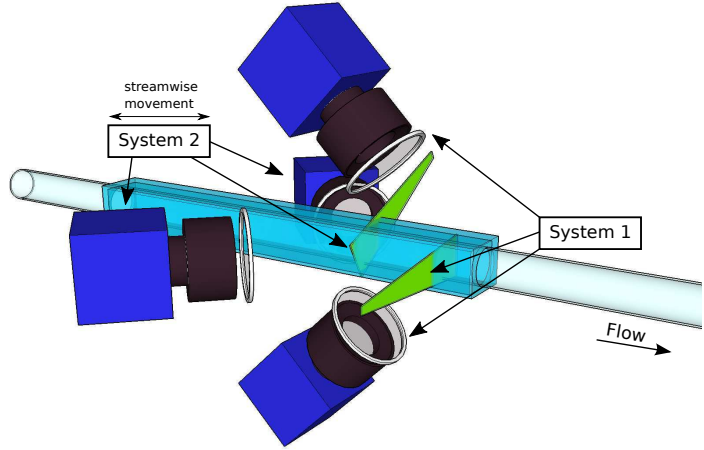}
	\end{center}
\caption{The dual-plane stereo PIV experimental setup. The two systems are separated by orthogonally polarizing the two laser sheets and mounting polarizing filters to each camera. System 2 is mounted on a traverse such that it can be moved in the streamwise direction.}
\label{fig:setup}
\end{figure}

The experiments were conducted in a $200 D$ long pipe facility, consisting of seven, $1.2$~m long, glass sections with an inner diameter $D = 38.1 \pm 0.025$~ mm. The velocity field was simultaneously acquired in two cross-sectional planes using two stereoscopic PIV systems (2D-3C), with an adjustable streamwise displacement, $\xi$. As shown in figure \ref{fig:setup}, the first PIV system was held fixed and consisted of a pair of 5.5 Megapixel LaVision Imager sCMOS cameras arranged vertically above and below the pipe. The second system consisted of a pair of 4.0 Megapixel LaVision Imager SX cameras mounted horizontally along with a LaVision articulating light delivery arm, and the  system was mounted on a low friction rail and driven by a traverse such that it could be moved as a unit in the streamwise direction. 

The working fluid was water seeded with $10~\mu$m glass hollow spheres.  The test section was enclosed by a rectangular acrylic box, filled with water to minimize the optical distortion due to refraction through the pipe wall, see figure \ref{fig:setup}. An access port was located immediately downstream of the test section in order to insert the stereo PIV calibration target while the pipe was filled with water. The target had 272 dots set in a rectangular grid, and was traversed $2$~mm in each direction of the laser sheets, resulting in three calibration images for each camera. 

A total of 21 data sets each containing 2000 image pairs were acquired for each Reynolds number. The streamwise distance between the two interrogation planes ($\xi/R$) was logarithmically varied, such that $(\xi/R) \in$ \{0.0; 0.0262; 0.0357; 0.0488; 0.0672; 0.0919; 0.126; 0.171; 0.234; 0.320; 0.438; 0.598; 0.818; 1.12; 1.53; 2.09; 2.86; 3.90; 5.34; 7.30; 9.97\}.  The separation distance was measured with a dial micrometer with a $12.7~\mu$m resolution, and to reduce the overall error it was based on the cumulative displacement referenced from the origin. Following the procedure of \cite{Hellstrom2015}, the two laser sheets were orthogonally polarized using two independent $l/2$-wave plates and each camera was subsequently equipped with linear polarizers such that they only identified particles situated within its own system. Three Reynolds numbers were investigated, $Re_\tau = u_\tau R/\nu \in \{1310;~2430;~3810\}$, or $Re_D = U_b(2R)/\nu \in \{51,000; 102,100; 168,700\}$, where $U_b$ is the bulk velocity.  The systems were operated at $f = 10$~Hz, corresponding to a convective bulk displacement of $U_b/(R f) \in \{7.05;~14.1;~23.3\}$ between snapshots, and all snapshots in the time series can therefore be considered statistically independent.
 
%------------------------------------------------------
% Methods ---------------------------------------------
%------------------------------------------------------
\section{Data Analysis}
\label{ch:method}

\begin{figure}
	\begin{center}
	\begin{tabular}{c c}
		\sidesubfloat[]{\includegraphics[height=.45\textwidth]{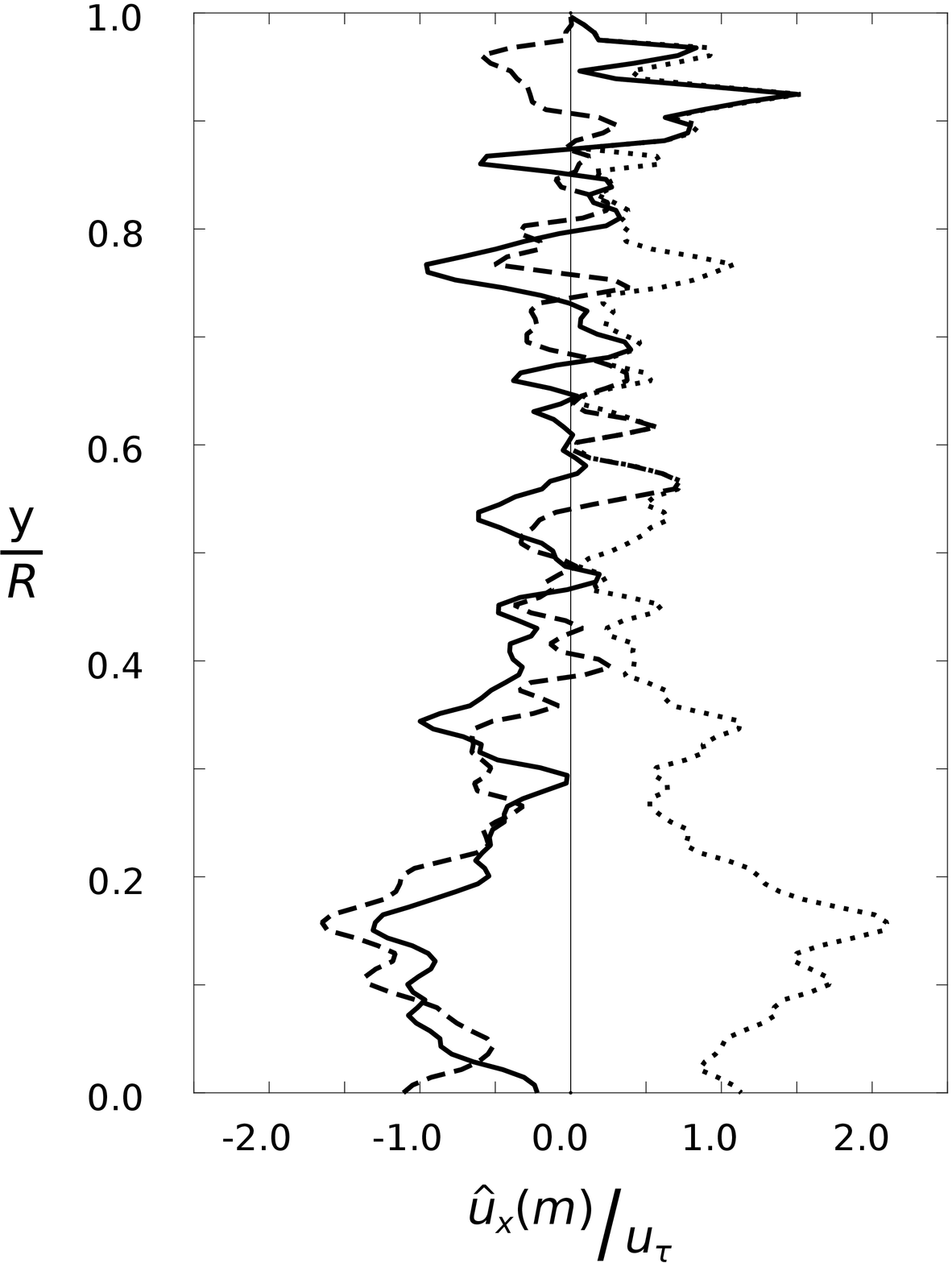}}&
		\sidesubfloat[]{\includegraphics[height=.45\textwidth]{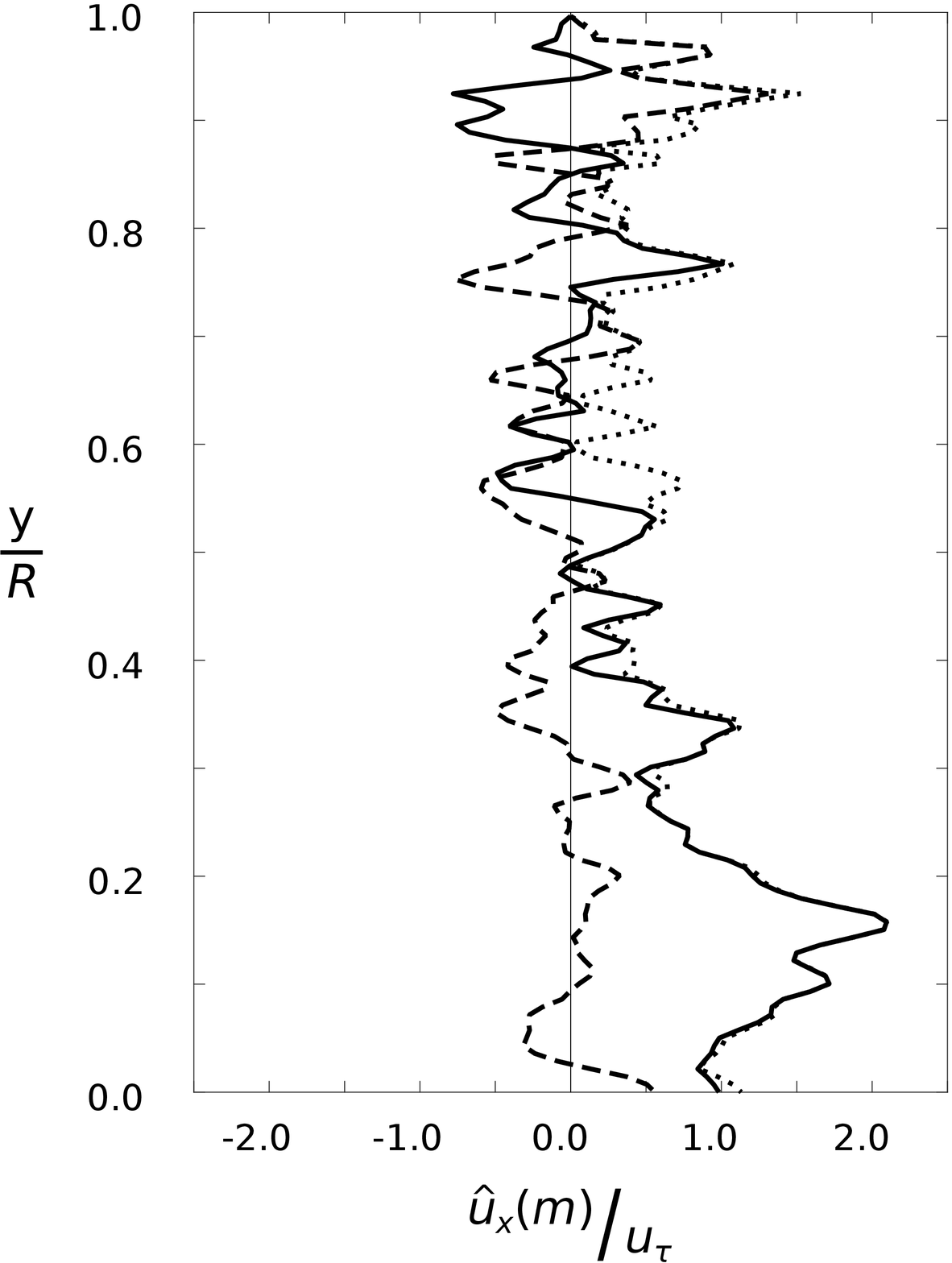}} 
	\end{tabular}
	\end{center}
\caption{
A representative, instantaneous streamwise velocity $(\hat{u}_x)$ profile for the fifth azimuthal mode $(m = 5)$ and $Re_\tau =2430$. (a) original velocity field, (b) the velocity field subject to a phase shift optimizing the real part of the profile. Real part --- ; imaginary part $- -$; absolute value of the profile $\cdot \cdot \cdot$.  Note that $y$ is measured from the wall, $y=R-r$.
}
\label{fig:instVel}
\end{figure}

We aim to identify a set of self-similar hierarchical structures which may be considered candidates for Townsend's attached eddy hypothesis. Previous work by \cite{Hwang2015} and \cite{Hellstrom2016a} showed that a spanwise Fourier decomposition of the fluctuating velocity field was a convenient initial step, and so we start there. This step acts as a sorting (or binning) procedure, where all structures with the same spanwise width are grouped together and represented by their azimuthal Fourier mode number, $m$, indicative of the number of azimuthally aligned structures at each instance. Figure \ref{fig:instVel}(a) shows a representative instantaneous velocity profile for $m = 5$ at $Re_\tau = 2430$, where $y$ is measured from the wall ($y=R-r$). The azimuthal orientation of an instantaneous structure with respect to our $\theta =0$ coordinate can be found from the phase of the complex velocity profile.  At the particular instance shown in figure \ref{fig:instVel}(a), we can see that both the real and imaginary components of the profile are negative, and of similar magnitude.

We will now derive a representative average eddy velocity profile for each mode $m$, represented by $u_\phi(m;r)$. \cite{Hellstrom2016a} and \cite{Hellstrom2017Conf} performed a proper orthogonal decomposition in the radial direction and showed that the first few POD modes were self-similar. Here, we take an alternative approach and search for an average eddy velocity profile, $u_\phi$, which, as a concept, lies closer to Townsend's hypothesis and also does not suffer from any truncation restrictions.  As shown by \cite{Adrian2000}, the large scale motions come in a pseudo aligned manner, where areas of low-momentum fluid are followed by areas of high-momentum fluid, such that their long-time average goes to zero with respect to the mean. Hence, we cannot simply average the Fourier decomposed velocity field with respect to time, as it would average to zero, that is, $\langle \hat{u}(m;r,t)\rangle_t = 0$. This fundamental feature holds for the current data-set; that is, there is no preferred azimuthal alignment for the eddies and their azimuthal distribution can, for all purposes here, be considered random.  To create an average eddy velocity profile, $u_\phi$, we will instead artificially align all structures of the same size before computing the average.  We achieve this alignment by performing an azimuthal phase shift ($\omega$) on each instantaneous velocity field, for each azimuthal mode number $(m)$, such that its phase is 0. However, the phase for an azimuthally leaning structure changes with wall-normal distance and we need to consider a representative phase. From an algorithmic perspective, we therefore maximize the real part of the streamwise velocity profile in the $l^2$-norm sense by evaluating

\begin{equation}
\max_\omega \frac{\int_0^R \left(\Re(\hat{u}_x(m;r,t)e^{i \omega m})\right)^2 r dr}{\int_0^R \left(\Im(\hat{u}_x(m;r,t)e^{i \omega m})\right)^2 r dr}.
\label{eq:phaseOpt}
\end{equation}

The phase shift procedure is illustrated in figure \ref{fig:instVel}, where \ref{fig:instVel}(a) is the original instantaneous velocity profile and \ref{fig:instVel}(b) has been azimuthally shifted. The procedure considers and rotates each snapshot and azimuthal Fourier mode separately, and does not take account of any interaction between snapshots and scales.  However, as each snapshot is taken to be statistically independent and Townsend's attached eddy hypothesis does not consider any scale interactions (or non-linearities), these assumptions are in line with the generic approach taken in this work. One advantage with this procedure, compared to considering the absolute value of each instantaneous velocity field, is that the turbulence noise will average out instead of being accumulated.
\begin{figure}
 	\begin{center}
 	\begin{tabular}{c c c}
		\sidesubfloat[]{\includegraphics[width=0.28\textwidth]{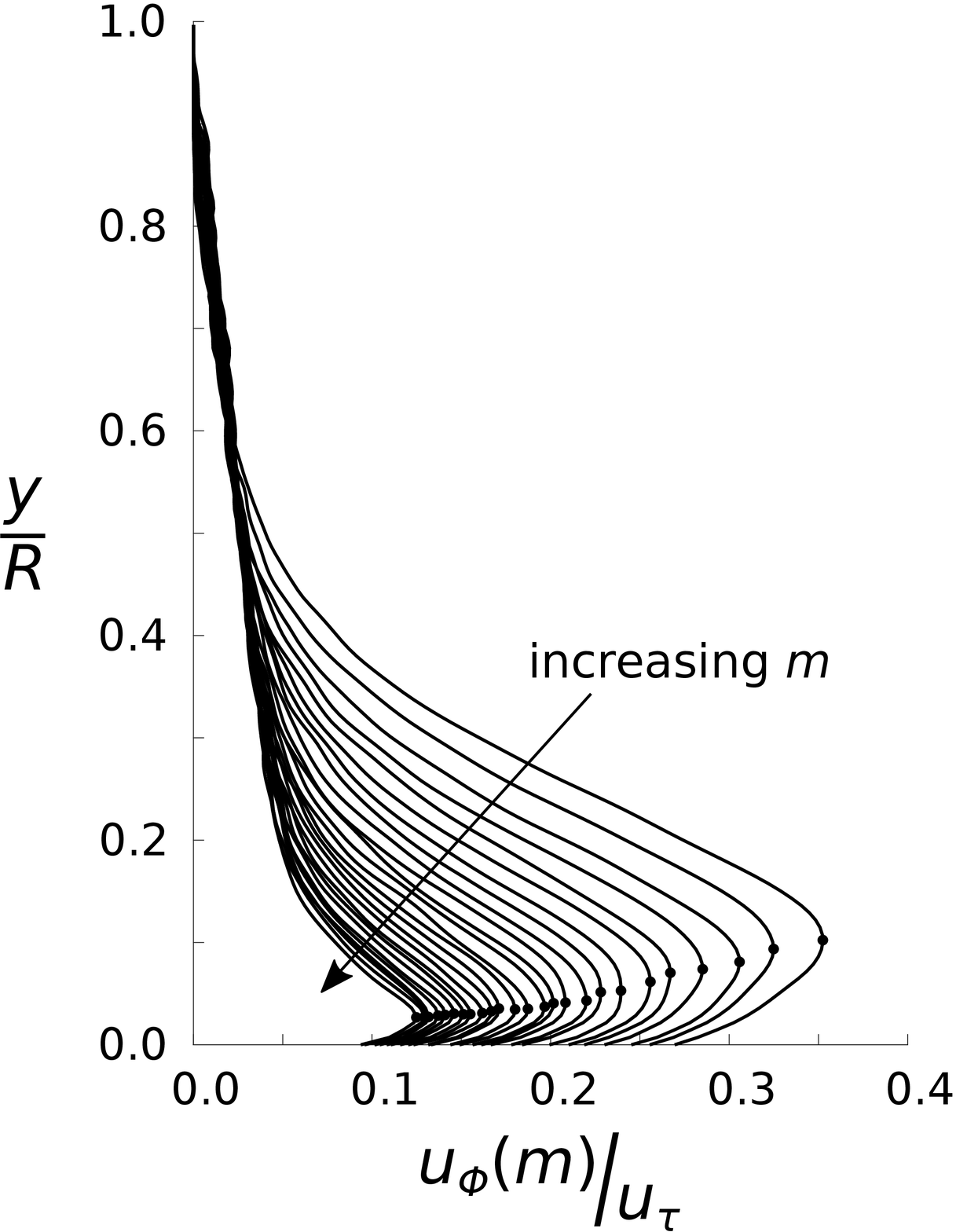}} &
		\sidesubfloat[]{\includegraphics[width=0.28\textwidth]{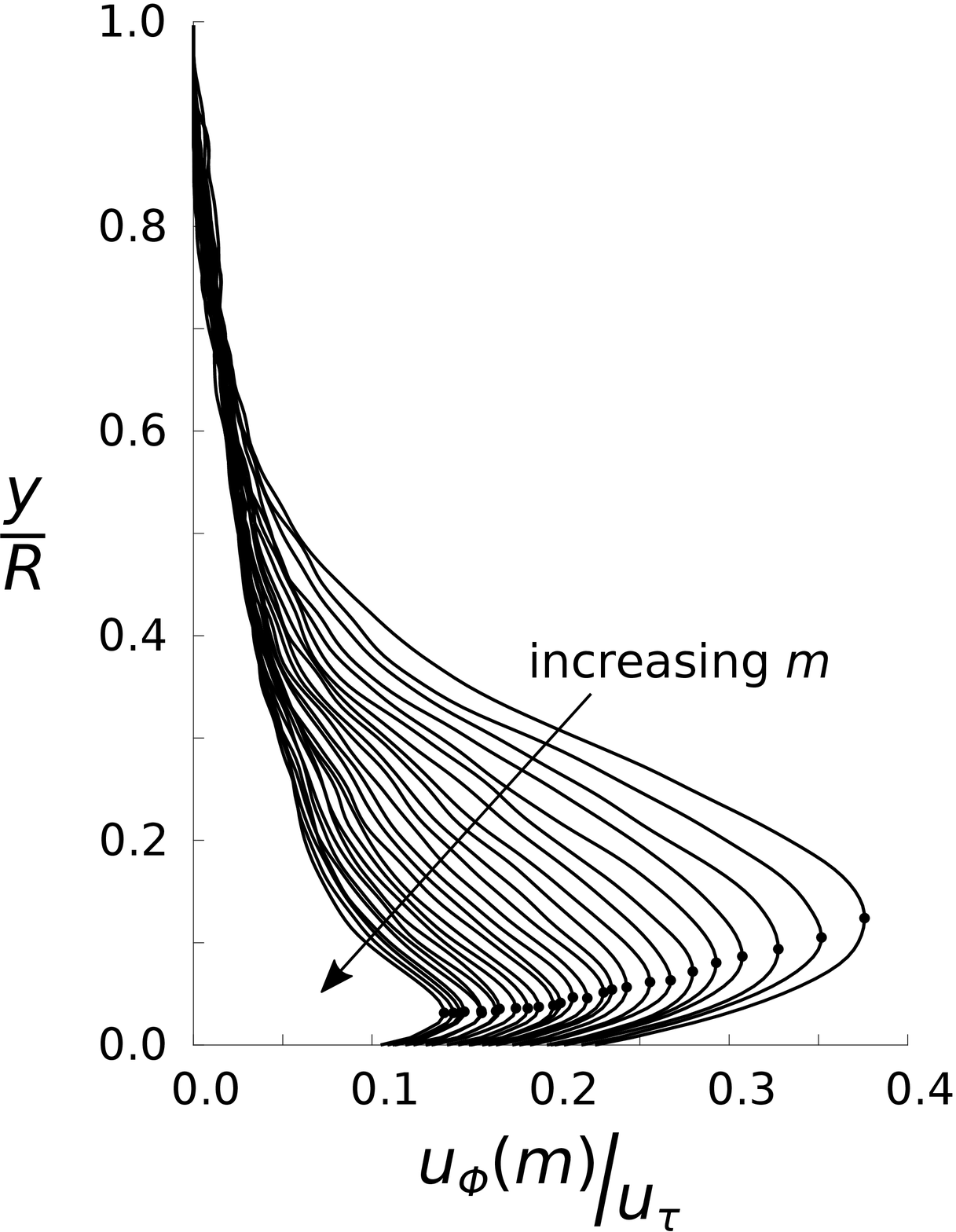}} &
		\sidesubfloat[]{\includegraphics[width=0.28\textwidth]{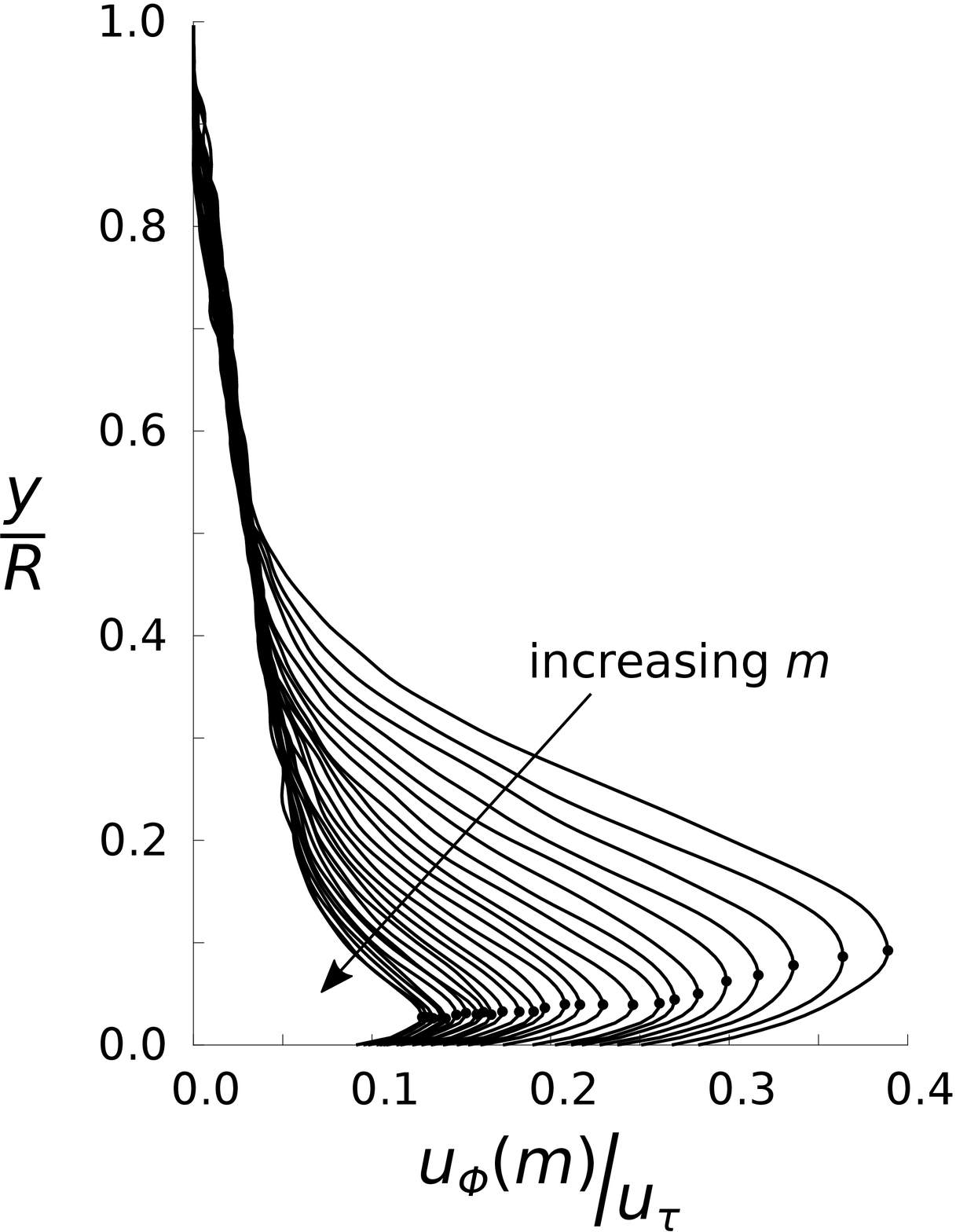}}\\
	\end{tabular}
 	\end{center}
 	\caption{The average eddy velocity profiles for all $m \in [10, 35]$. (a) $Re_\tau = 1310$. (b) $Re_\tau = 2430$. (c) $Re_\tau = 3810$. The symbol $\bullet$ marks the peak location for each eddy velocity profile. }
\label{fig:velModes}
\end{figure}
This phase shift was applied to each instantaneous velocity field in the complete data set, for all azimuthal modes $m$. The representative eddy velocity profile $u_\phi(m;r)$ was obtained by performing a temporal average of the  rectified velocity fields, and the results are shown in figure \ref{fig:velModes}. In order to assess the robustness of the phase alignment process the same procedure was performed while subjecting equation \ref{eq:phaseOpt} to some modifications, namely, i) introducing an adaptive integration limit $r\in(0, \lambda_\theta/2)$ and ii) removing the radial weighting $r$ in the $l^2$ integration. Neither of these changes imposed any noticeable changes to the peak location or the magnitude of the eddy velocity profiles. The resulting eddy velocity profiles can be seen in figure \ref{fig:velModes}, where the change in profile can be seen as a function of azumuthal mode number ($m$). 

When considering an averaging process, as is done in this work, it is also essential to understand what an average represents, and we still need to verify that the averaging procedure that generated the eddy velocity profiles is meaningful. Because how the structures are sorted by their spanwise width, it is crucial to inspect the repetition rate of the structures within each bin. Consider a hypothetical set of self-similar structures that collapses completely when scaled by their height and a common velocity scale (the friction velocity $u_\tau$).  If, however, their probability to occur decreases with their size, then the no-activity space between consecutive structures of a given size increases as their size decreases. This would lead to probability density functions (PDFs) with more than one peak, representing the active and inactive zones, respectively.  As an illustration, consider an $m=5$ structure with a velocity strength of 10$u_\tau$ that is present for a distance $1R$ and absent for a distance $1R$. Its average velocity strength would be 5$u_\tau$. Now consider an $m=10$ structure, with the same velocity strength 10$u_\tau$ (that is, it has the same velocity scale).  If this $m=10$ structure is present for $0.5R$ and absent for $0.5R$, its average would also be 5$u_\tau$. However, we only know that the structure should be geometrically self-similar, and we can make no prediction about how often it occurs.  If it were present for  $0.5R$ and absent for $4.5R$, the average strength would be $u_\tau$, and it would incorrectly appear as a weaker structure.  Thus the averaging process used to estimate their strength (that is, their velocity scale) must also account for their probability of occurrence.

\begin{figure}
 	\begin{center}
		\begin{tabular}{c c c}
			\sidesubfloat[]{\includegraphics[width=.29\textwidth]{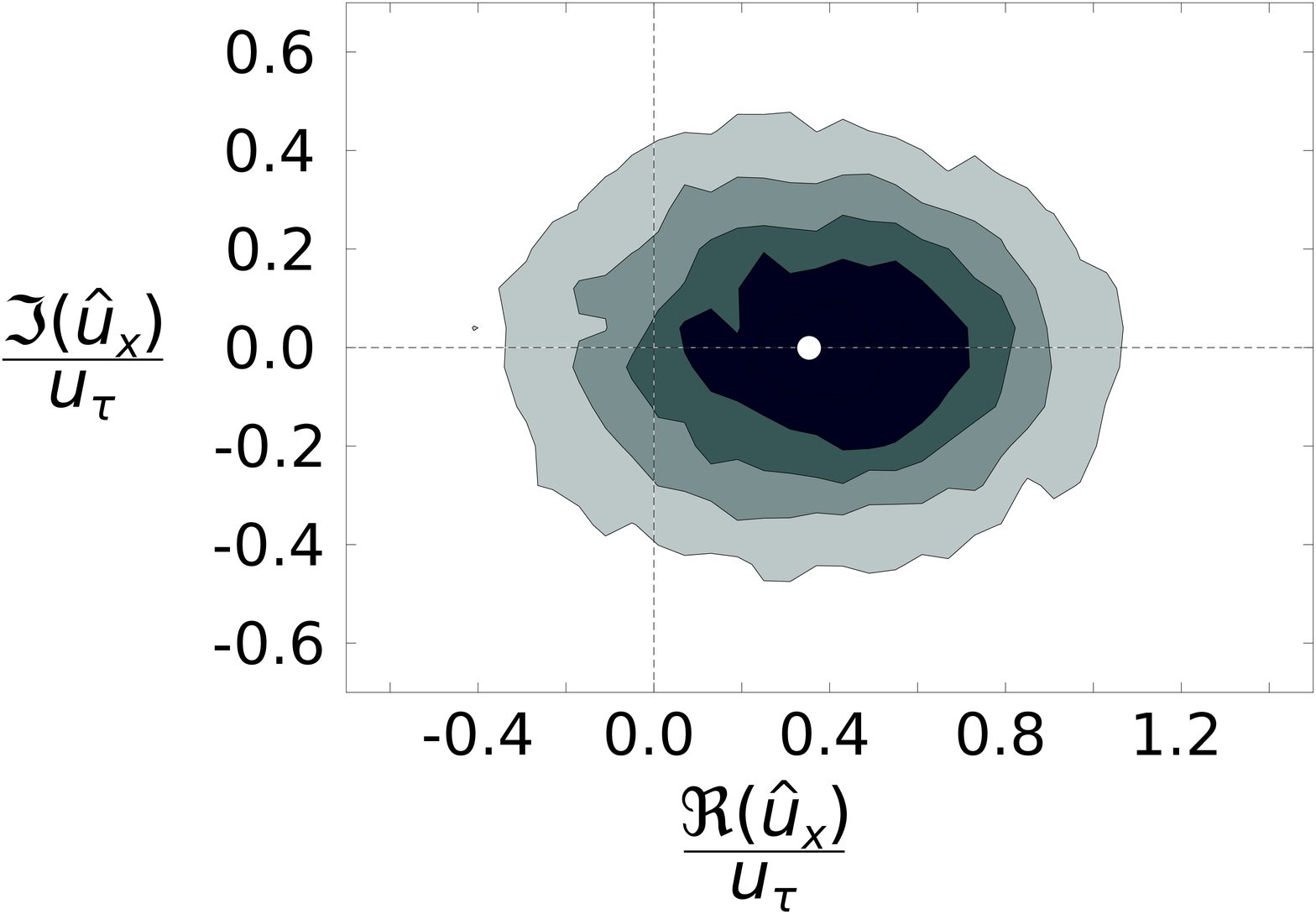}} &
			\sidesubfloat[]{\includegraphics[width=.29\textwidth]{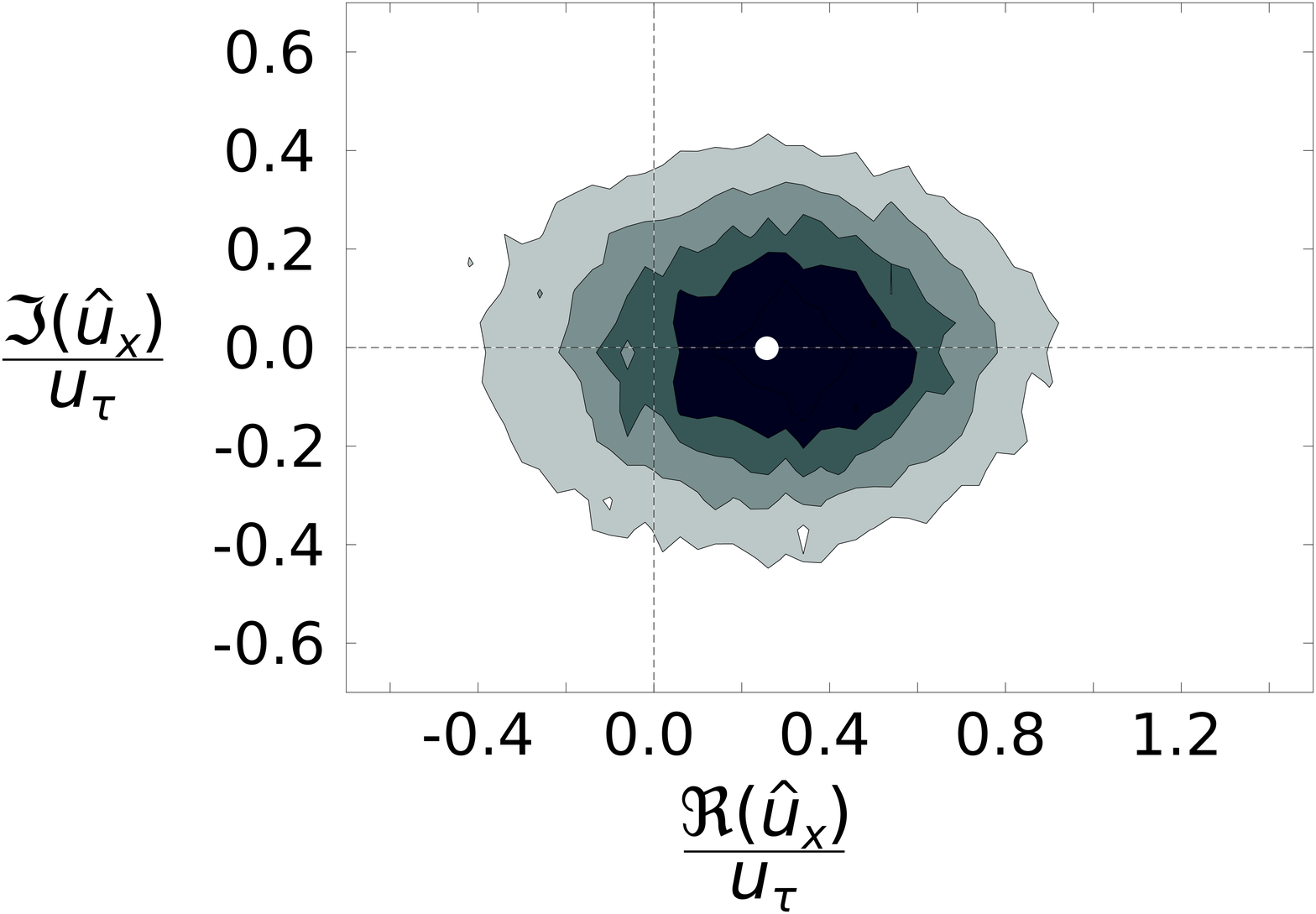}} &
			\sidesubfloat[]{\includegraphics[width=.29\textwidth]{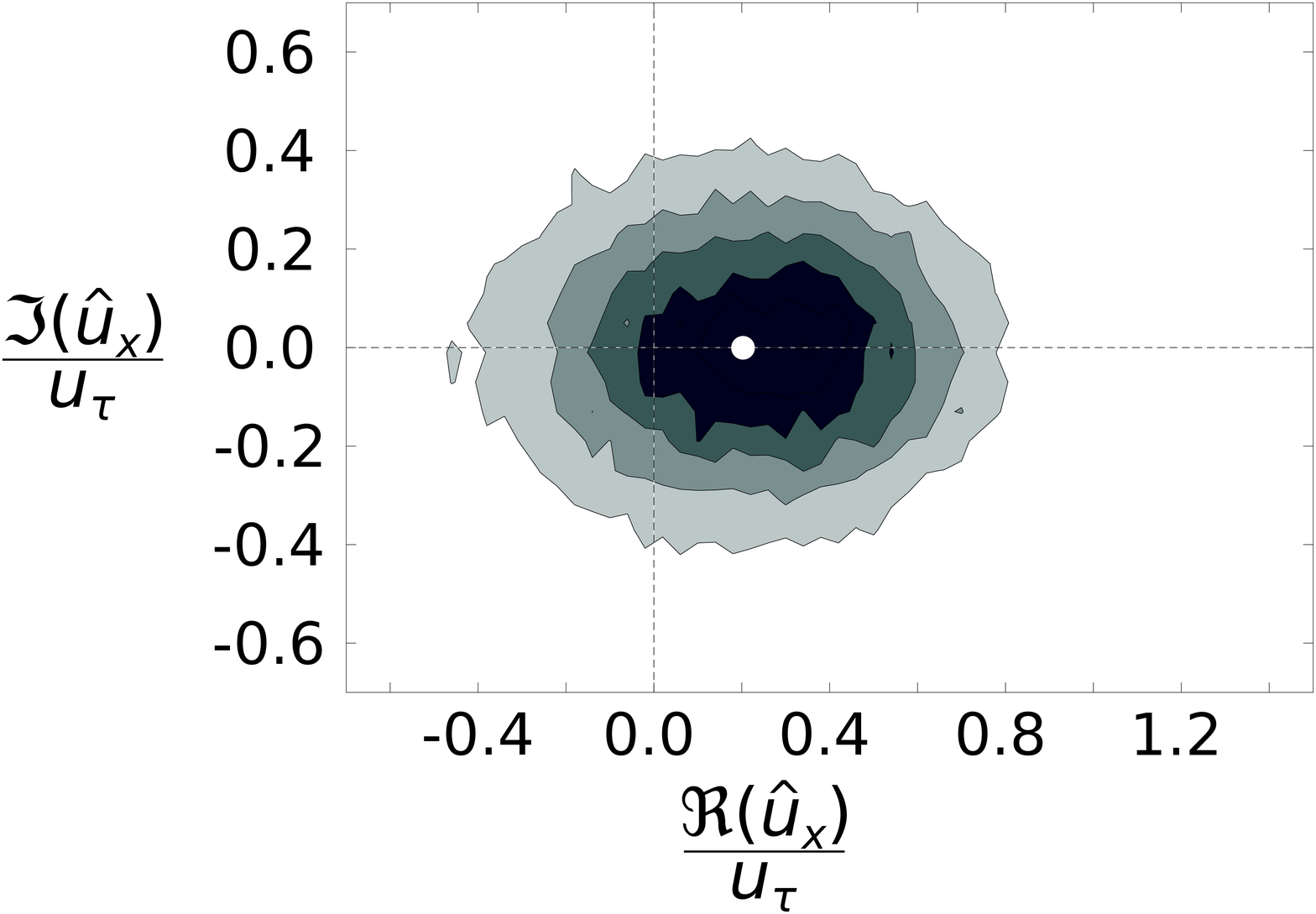}} \\
			\sidesubfloat[]{\includegraphics[width=.29\textwidth]{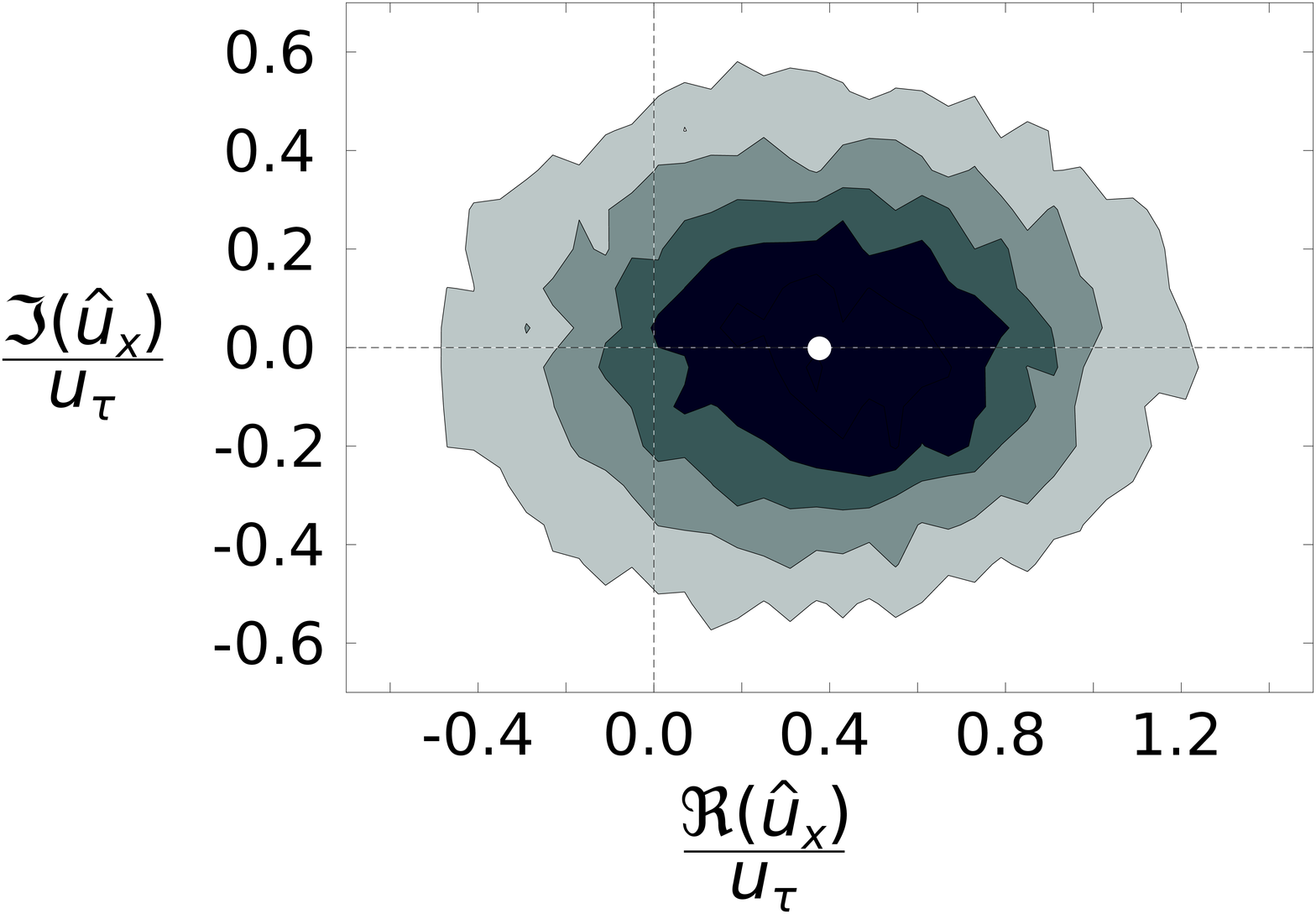}} &
			\sidesubfloat[]{\includegraphics[width=.29\textwidth]{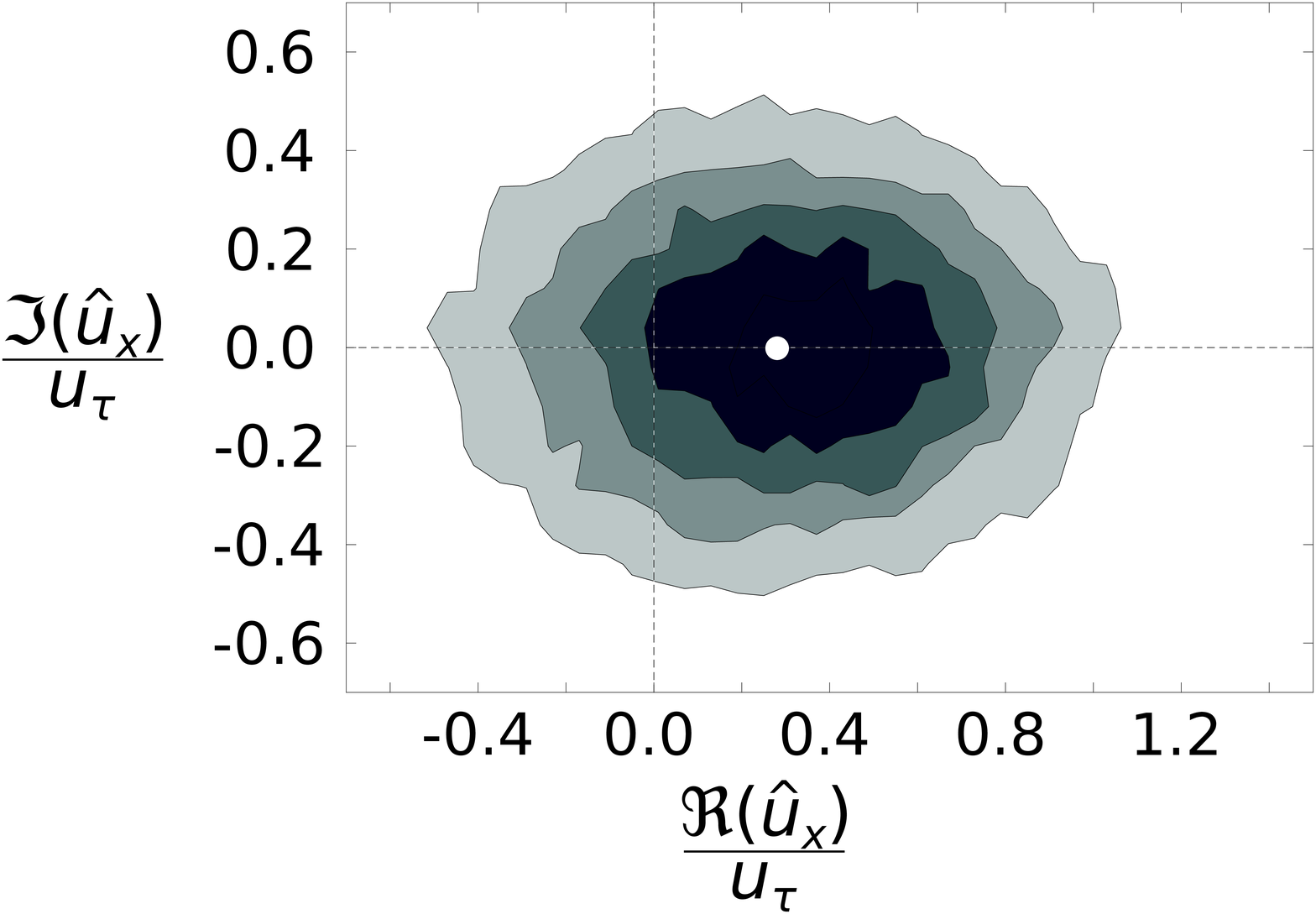}} &
			\sidesubfloat[]{\includegraphics[width=.29\textwidth]{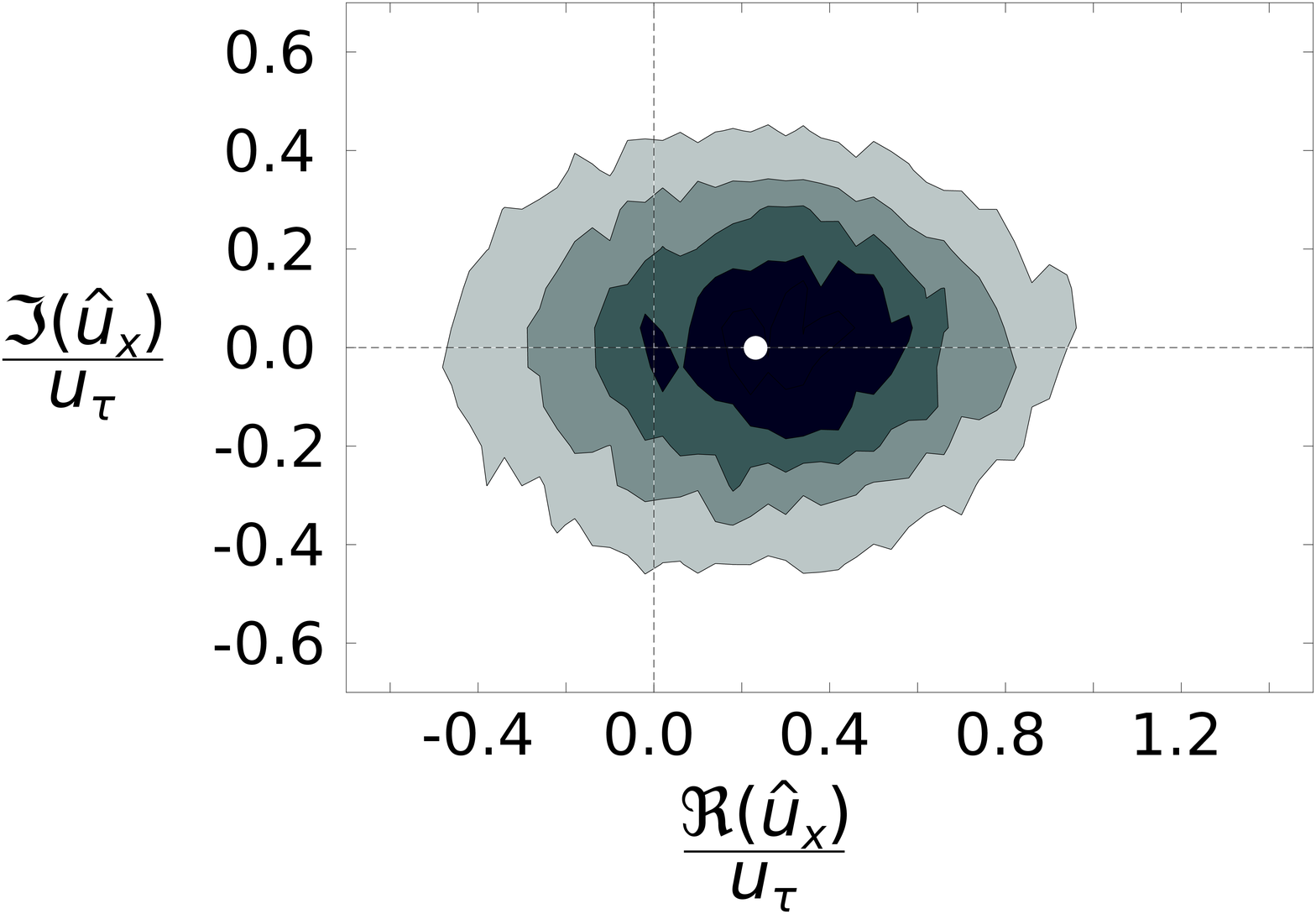}} \\
			\sidesubfloat[]{\includegraphics[width=.29\textwidth]{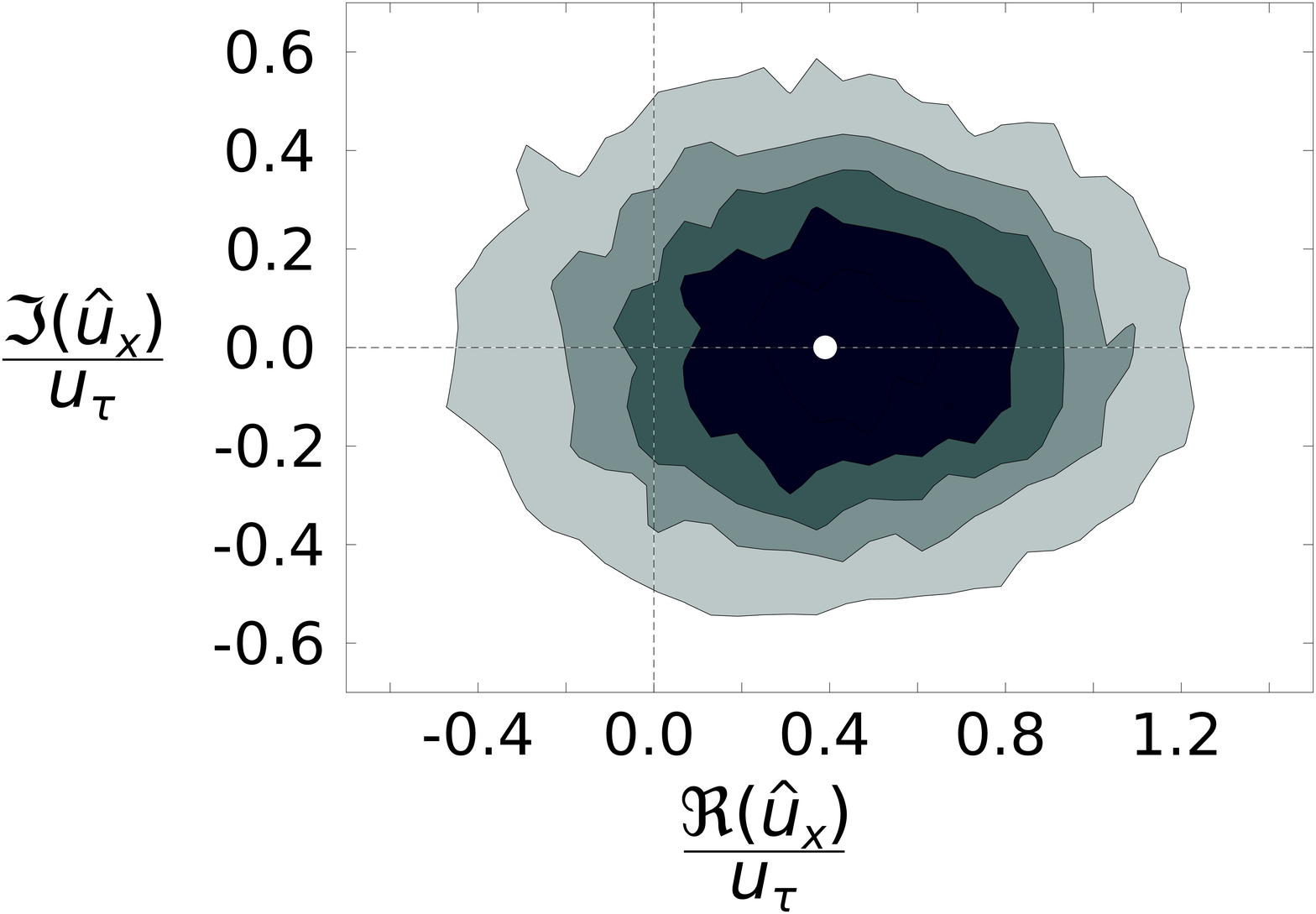}} &
			\sidesubfloat[]{\includegraphics[width=.29\textwidth]{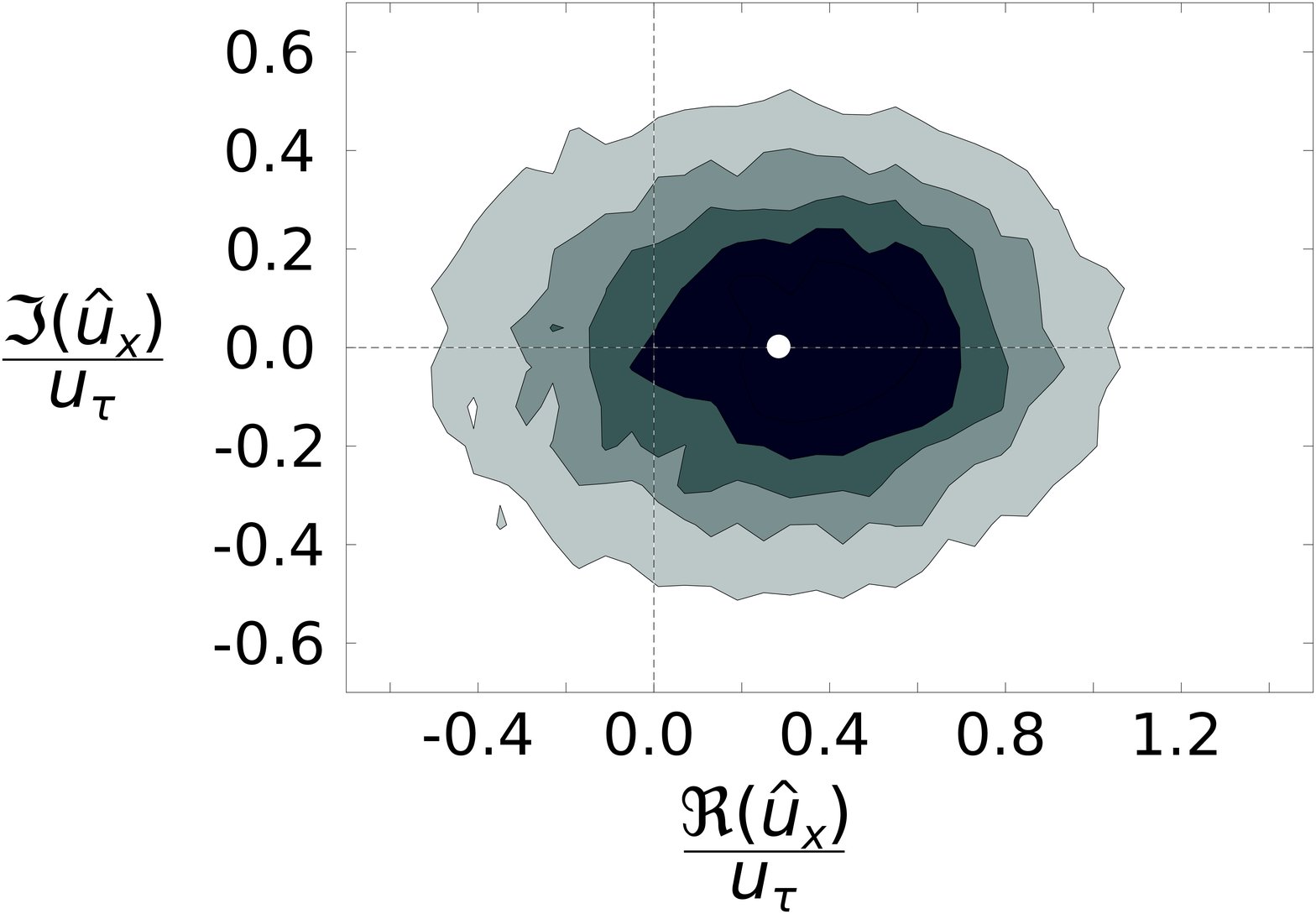}} &
			\sidesubfloat[]{\includegraphics[width=.29\textwidth]{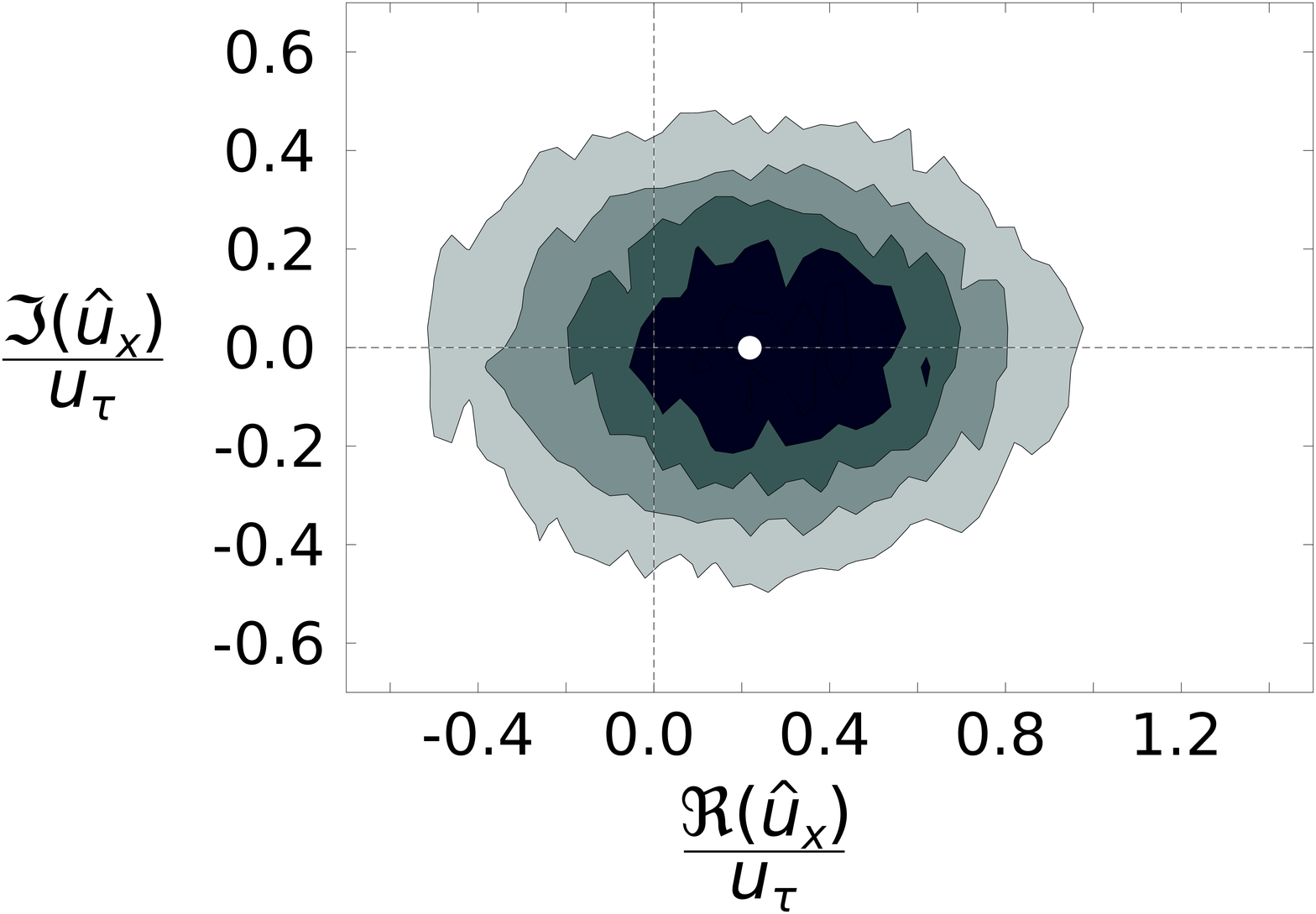}}
		\end{tabular}
	\end{center}
\caption{PDFs of the streamwise velocity component $u_x$, evaluated at the peak eddy velocity location found in figure \ref{fig:velModes} (real and imaginary parts). The corresponding peak eddy velocity found in figure \ref{fig:velModes} is shown with a solid white circle. (a) $Re_\tau = 1310$, $m = 10$. (b) $Re_\tau = 1310$, $m = 15$. (c) $Re_\tau = 1310$, $m = 20$. (d) $Re_\tau = 2430$, $m = 10$. (e) $Re_\tau = 2430$, $m = 15$. (f) $Re_\tau = 2430$, $m = 20$. (g) $Re_\tau = 3810$, $m = 10$. (h) $Re_\tau = 3810$, $m = 15$. (i) $Re_\tau = 3810$, $m = 20$. Each PDF is scaled by its peak value and is shown with contour levels \{0.2, 0.4, 0.6, 0.8\}.}
\label{fig:uPDF}
\end{figure}

Consequently, we need to verify the validity of our averaging procedure by inspecting the velocity PDF for each structure size. Figure \ref{fig:uPDF} shows the PDFs for three different structure sizes, $m \in\{10,~15,~20\}$, for all three Reynolds numbers. For consistency, each PDF is evaluated at the eddy peak location as found in figure \ref{fig:velModes}. We consider the Fourier transformed velocity field, which is complex, and the PDF is visualized in the complex plane. The corresponding eddy peak velocity found in figure \ref{fig:velModes} is shown with a solid white circle, and it consistently falls in the vicinity of the PDF peak.  There is only one peak, and so all eddies of the same width will from here on be represented by their average profiles, as shown in figure \ref{fig:velModes}.

%------------------------------------------------------
% Results and discussion --- temporal comparison ------
%------------------------------------------------------
\section{Eddy self-similarity}
\label{ch:selfsim}
We start off by considering the azimuthal direction. As previously indicated, the azimuthal direction is already a self-similar dimension, that is, all structures are (on the basis of the Fourier decomposition) forced sine-waves. When decomposing a flow using POD, the basis functions are optimal in that the average turbulent kinetic energy of the first $n$ modes is maximized. If a direction is statistically homogeneous, and periodic (or infinite) the POD modes are reduced to a Fourier series (or Fourier transform). 
As a consequence of the axisymmetry of the pipe, the azimuthal Fourier modes are therefore the optimal decomposition, in a POD sense, in this direction. 
We will build on this observation and assume that the eddy width can be represented by the azimuthal wavelength, and that this wavelength can be used as the characteristic length scale.
The eddy width can then be defined as $\lambda_\theta = 2 \pi (R-y_{MAX}) /m$, where $y_{MAX}$ is the wall-normal location where the eddy velocity profile has its maximum, as seen in figure \ref{fig:velModes}.

\begin{figure}
 	\begin{center}
		\begin{tabular}{c c}
			\sidesubfloat[]{\includegraphics[height=.5\textwidth]{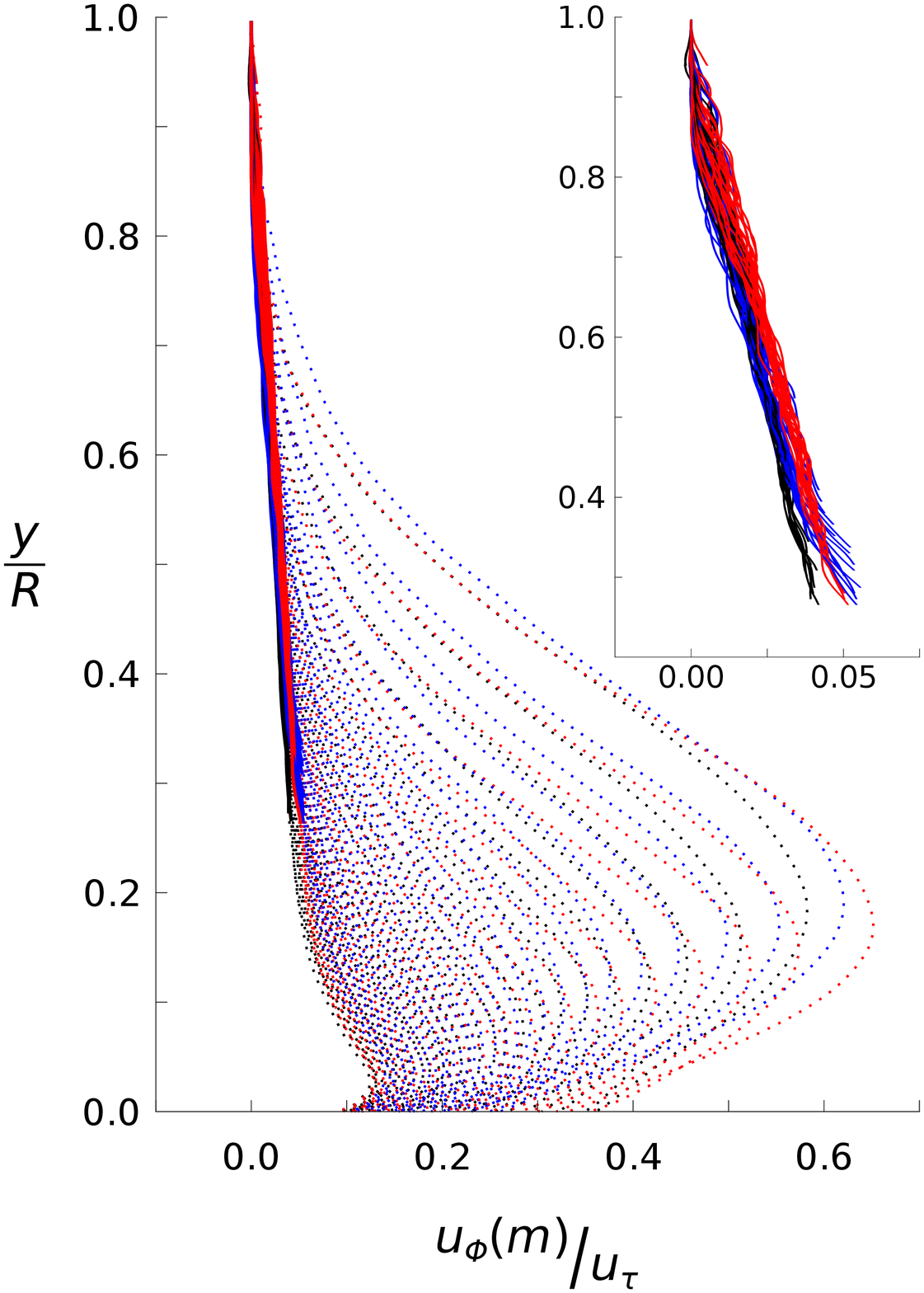}} &
			\sidesubfloat[]{\includegraphics[height=.5\textwidth]{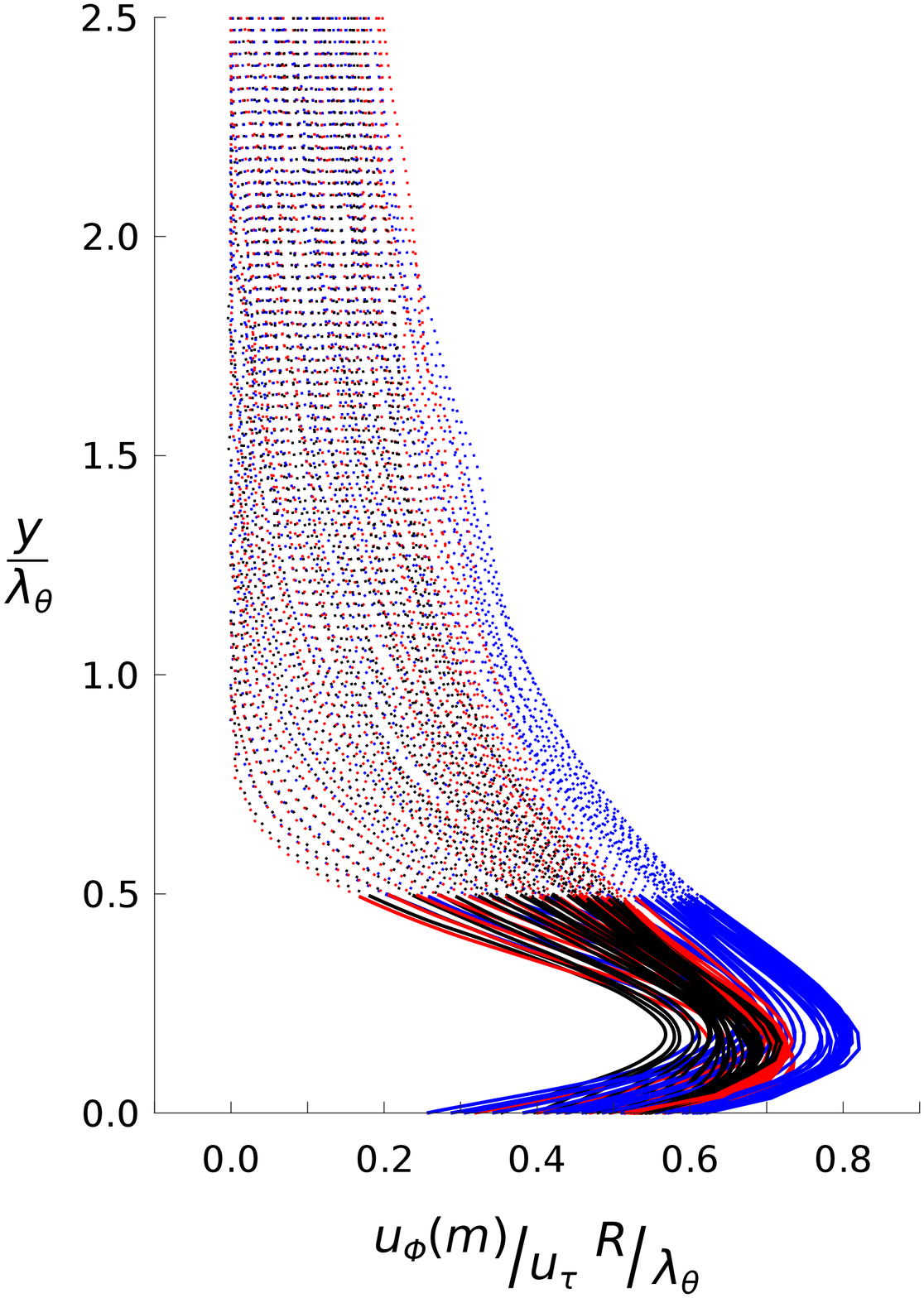}}				        
		\end{tabular}
	\end{center}
\caption{Eddy velocity profiles for all $m \in[5,35]$. Profiles for $Re_\tau = \{1310,~2430,~3810\}$ are shown in black, blue, and red line plots, respectively. (a) profiles are scaled with outer coordinates, the outer section of each profile ($y \geq 1.5\lambda_\theta$) is highlighted with a solid line. The insert shows a zoomed view of the outer section of the structures, and highlights the close collapse in this region. (b) profiles are scaled with eddy-size coordinates, the inner section of each profile ($y \leq 0.5\lambda_\theta$) is highlighted with a solid line.}
\label{fig:velModeScaled}
\end{figure}

Figure \ref{fig:velModeScaled}(a,b) show the averaged eddy profiles scaled in outer and eddy-size coordinates, respectively. Here, the eddy profiles are shown for the range $m \in [5,~35]$, or alternatively expressed as: $\lambda_\theta/R \in [1.03,~0.175]$ for $Re_\tau = 1310$; $\lambda_\theta/R \in [1.02,~0.174]$ for $Re_\tau = 2430$; and finally $\lambda_\theta/R \in [1.05,~0.175]$ for $Re_\tau = 3810$.

For the purpose of defining a cutoff between inner and outer regions, consider simple spherical eddies, where each positive eddy is flanked by two negative eddies (and vice versa). The azimuthal wavelength will span the width of a positive and a negative structure combined, and their characteristic height will consequently be $0.5\lambda_\theta$. We use this general guideline to define the outer region of the eddies as 3 times the height of the eddy or $1.5\lambda_\theta$ so that the near-wall behavior would not influence the outer flow behavior. In figure \ref{fig:velModeScaled}(a) the outer section of each eddy is highlighted with a solid line while the inner region is dotted, and we can see a complete similarity collapse for all profiles and Reynolds numbers in this region.  The cutoff determines where to begin the shaded region, and any value greater than two eddy heights, that is, greater than $\lambda_\theta$, would suffice.  

The inner region should, according to Townsend, be geometrically scaled by the eddy size, and in figure \ref{fig:velModeScaled}(b) we see a clear collapse in the wall-normal location of the peaks of all the eddy velocity profiles for all Reynolds numbers (located approximately at $y/\lambda_\theta=0.25$).  

However, it is more complicated to find a unified velocity scale. Townsend's model considers each eddy velocity as a contribution to a velocity deficit with respect to the potential flow solution (plug flow). 
According to Townsend's model, when integrating the contributions from all eddies, a constant mean stress profile is obtained, as well as the outer form of the logarithmic mean velocity profile (with respect to wall-distance) \citep{Perry1982}. In this work, we consider instead a fluctuating velocity field compared to its local mean value, and would therefore expect a somewhat different velocity scaling for the two different approaches. If we were to keep the local mean flow in our analysis, though, it would have a zero azimuthal periodicity and belong solely to the zeroth azimuthal Fourier mode ($m = 0$), and would therefore not account for the discrepancy between these two approaches. 
The two approaches also stress the importance of the characteristic convection velocity for each structure. For instance, \cite{Adrian2000} found that LSMs with different convective velocities have different characteristics.

However, when looking at figure \ref{fig:velModeScaled}(a) it is clear that the eddy velocity profiles require a scaling sensitive to eddy size. In figure \ref{fig:velModeScaled}(b) we have followed the argument from \cite{Perry1986} and assumed that the eddy velocity is proportional to  $yu_\tau$, or $u_\phi \propto y u_\tau \propto \lambda_\theta u_\tau$.

Although convincing, there is a noticeable spread between the scaled profiles.
It should be stressed that the magnitude spread of the collapsed velocity profiles is bound on the lower end by $Re_\tau = 1310$ and the upper end by $Re_\tau = 2430$, with $Re_\tau = 3810$ falling between, suggesting it is not a Reynolds number trend. Furthermore, within each data set, there is also no clear link to structure size. 
It is possible that the eddy velocity not only decreases with its size, but that it does so in a way that depends on its convection velocity, which is close to the local mean velocity. Alternatively, the deviations might also originate from the non-discriminatory averaging approach, which may include structures not driven by inertial forces and consequently could help dictate the eddy velocity profile. However, we will leave any further scaling attempts to future studies, and are satisfied with this demonstration of a preliminary collapse using a velocity scale that varies proportionally with $\lambda_\theta$, that is, wall distance $y$.

We have now achieved a self-similar eddy scaling for the azimuthal and wall-normal directions, and recognized a possible self-similar behaviour for the eddy velocity profiles. The fourth, and last, dimension is the streamwise length of each eddy, which will be examined using two-point correlations between the two PIV interrogation planes. The correlation map considers the Fourier transformed velocity fields, with one fixed reference point located at $y_{MAX}$. That is, we define
\begin{equation}
\rho(\eta,m,\xi) = \frac{\frac{1}{T} \sum_{t=1}^T \hat{u}_x(y_{MAX},m,x_0,t) \hat{u}_x^*(y_{MAX}+\eta,m,x_0+\xi,t)}{ \parallel \hat{u}_x(y_{MAX},m,x_0,t) \parallel_t},
\label{eq:corr}
\end{equation}
where $\wedge$, $\ast$ and $\parallel \cdot \parallel_t$ represent the Fourier transformed velocity field, its complex conjugate and the $l^2$-norm with respect to time, respectively.  Here, $\eta$ and $\xi$ represent the wall-normal and streamwise spatial shifts in the two-point correlations, respectively. 
It should also be clear that the two velocity fields used in these correlation maps have not been subject to the azimuthal phase shift, as it would be an unnecessary procedure. The two planes capture the same structure in their views, and should consequently be subject to the same phase shift (or none at all).
\begin{figure}
 	\begin{center}
		\begin{tabular}{c c c}
			\sidesubfloat[]{\includegraphics[width=0.28\textwidth]{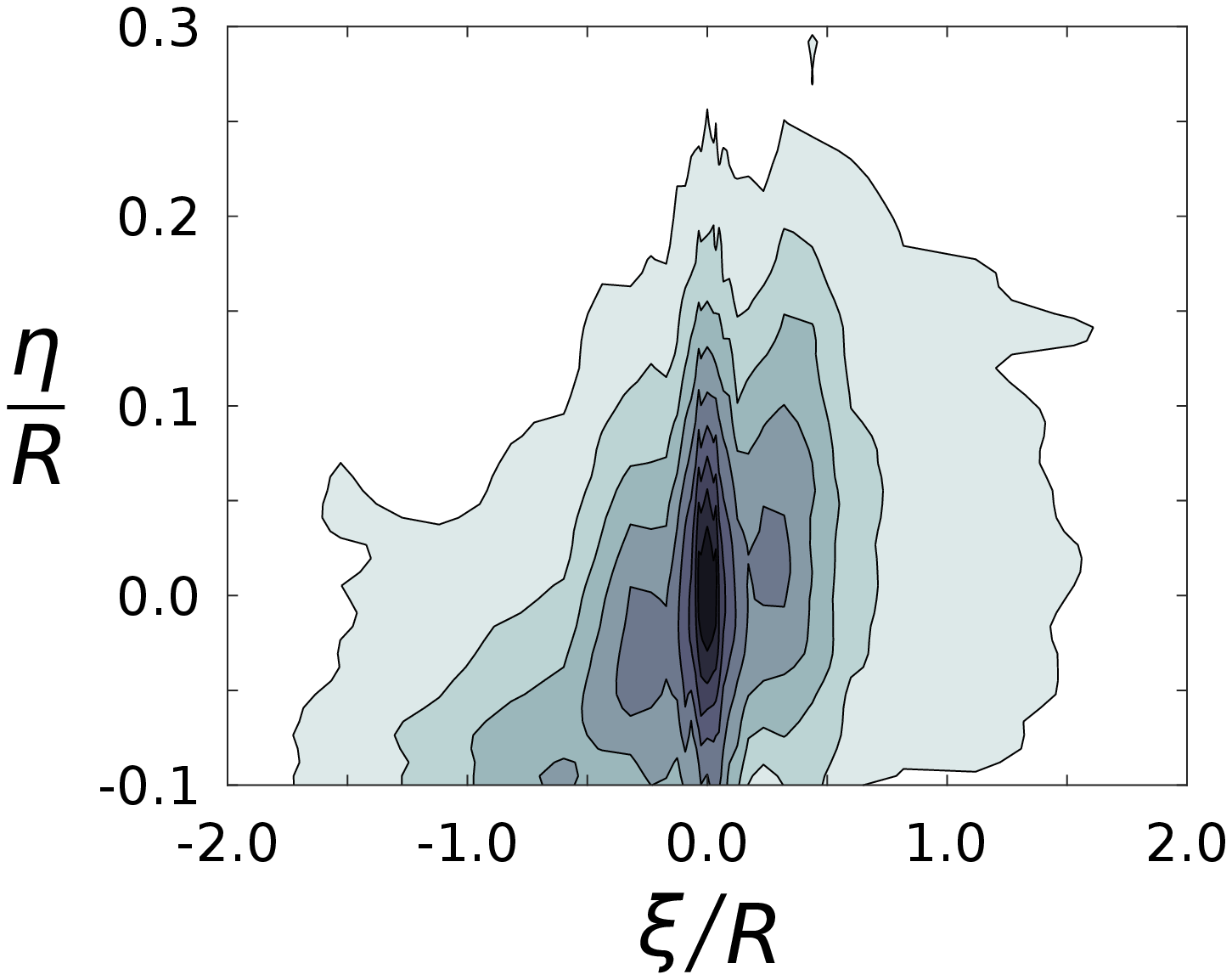}} &
			\sidesubfloat[]{\includegraphics[width=0.28\textwidth]{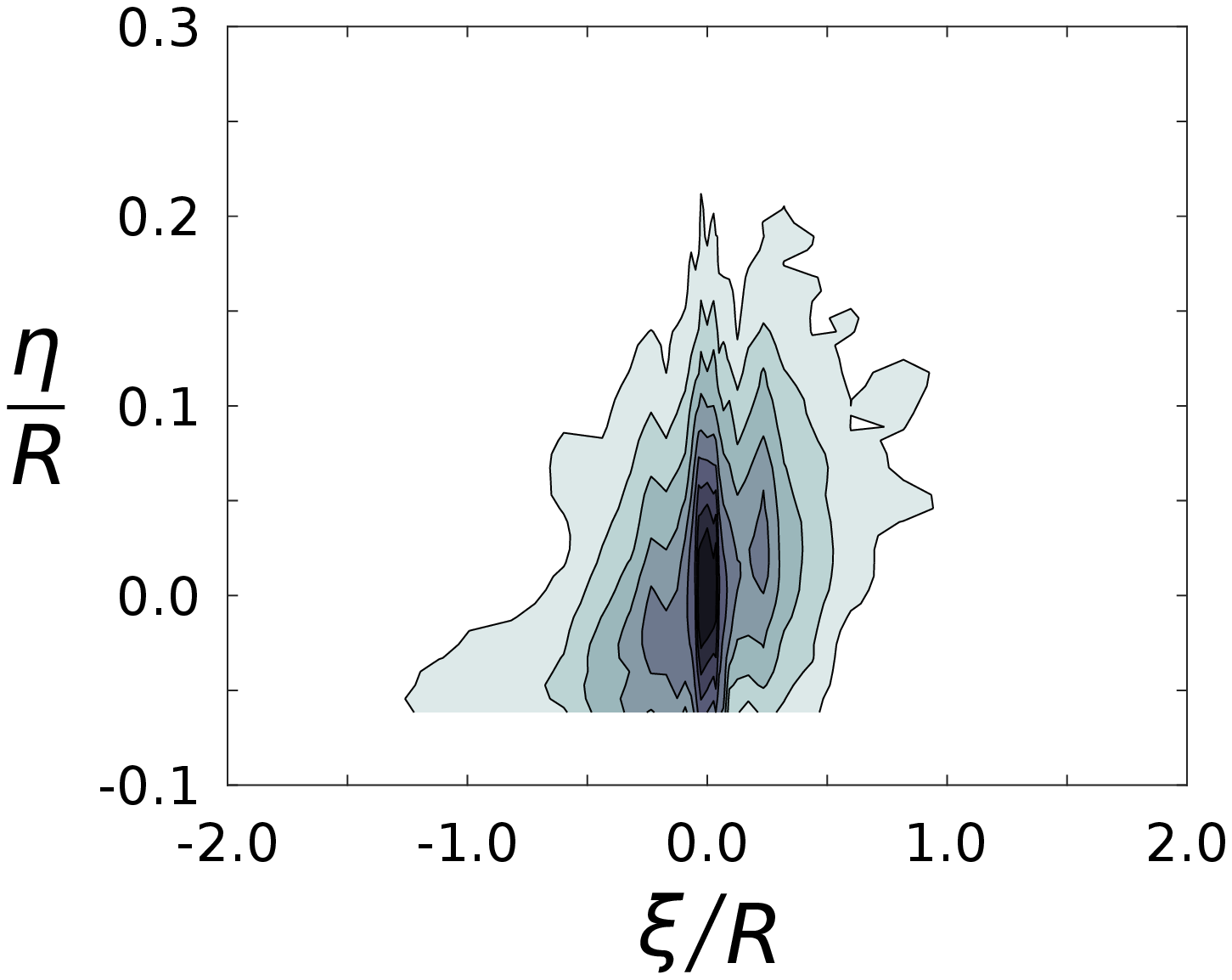}} &
			\sidesubfloat[]{\includegraphics[width=0.28\textwidth]{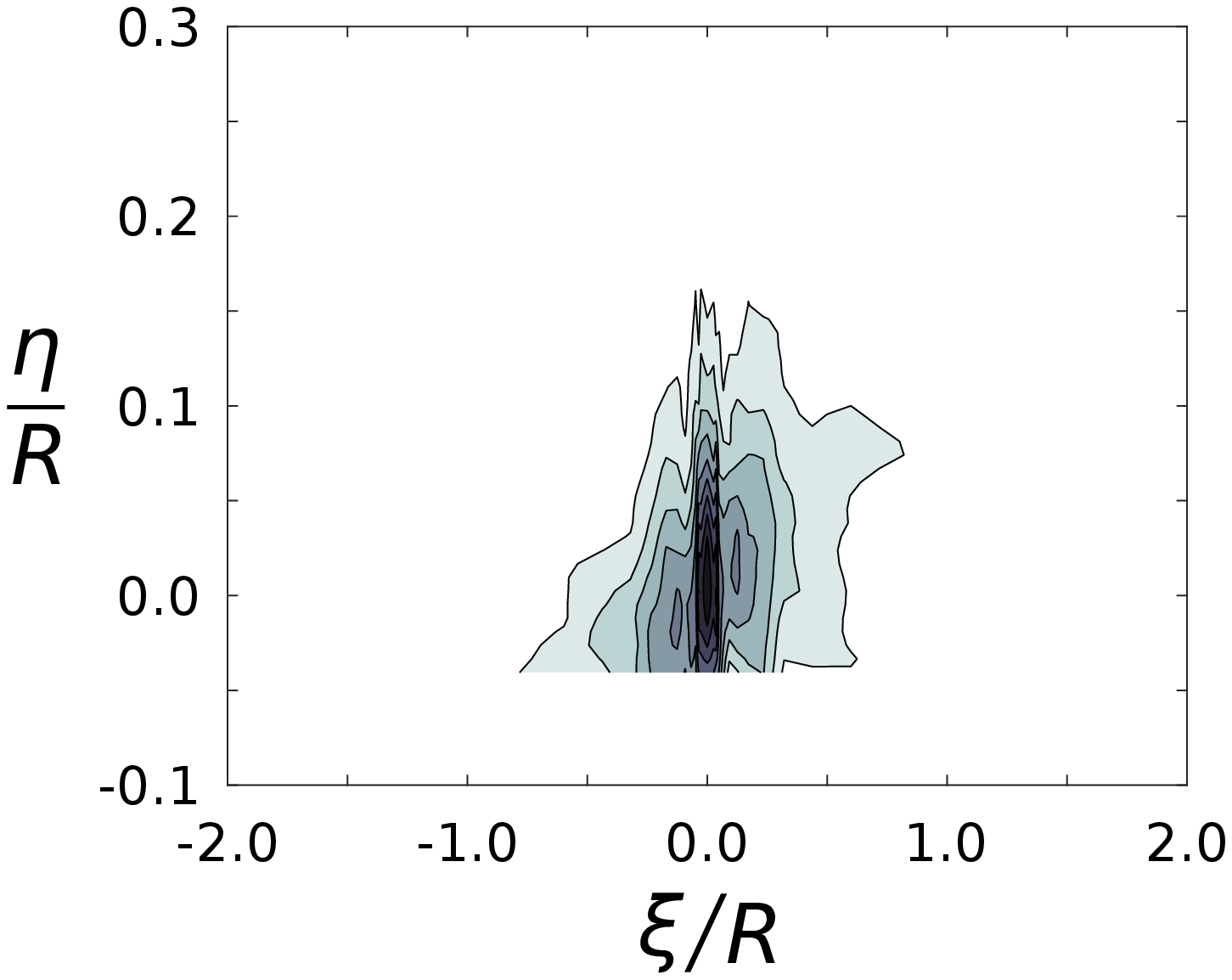}} \\
			\sidesubfloat[]{\includegraphics[width=0.28\textwidth]{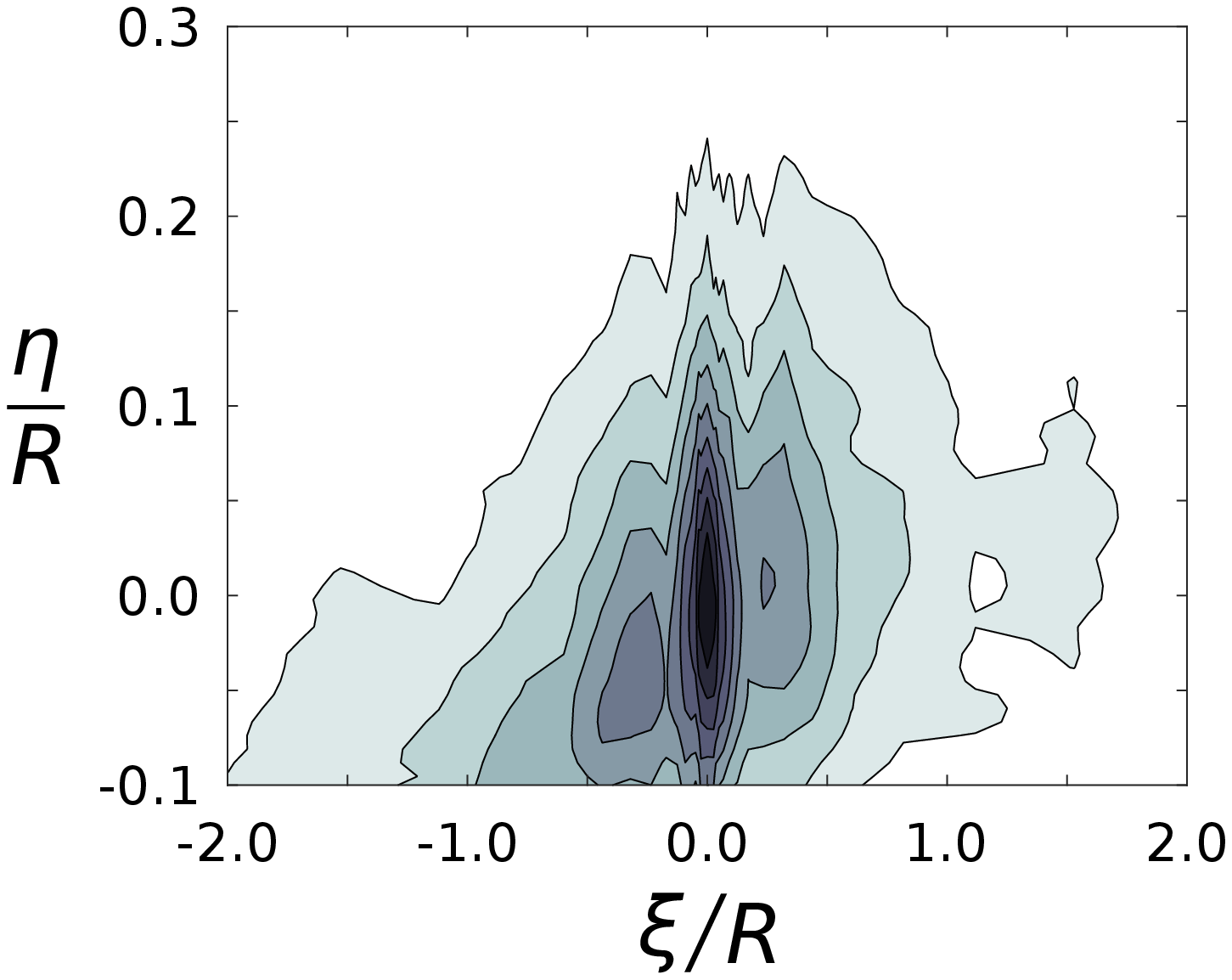}} &
			\sidesubfloat[]{\includegraphics[width=0.28\textwidth]{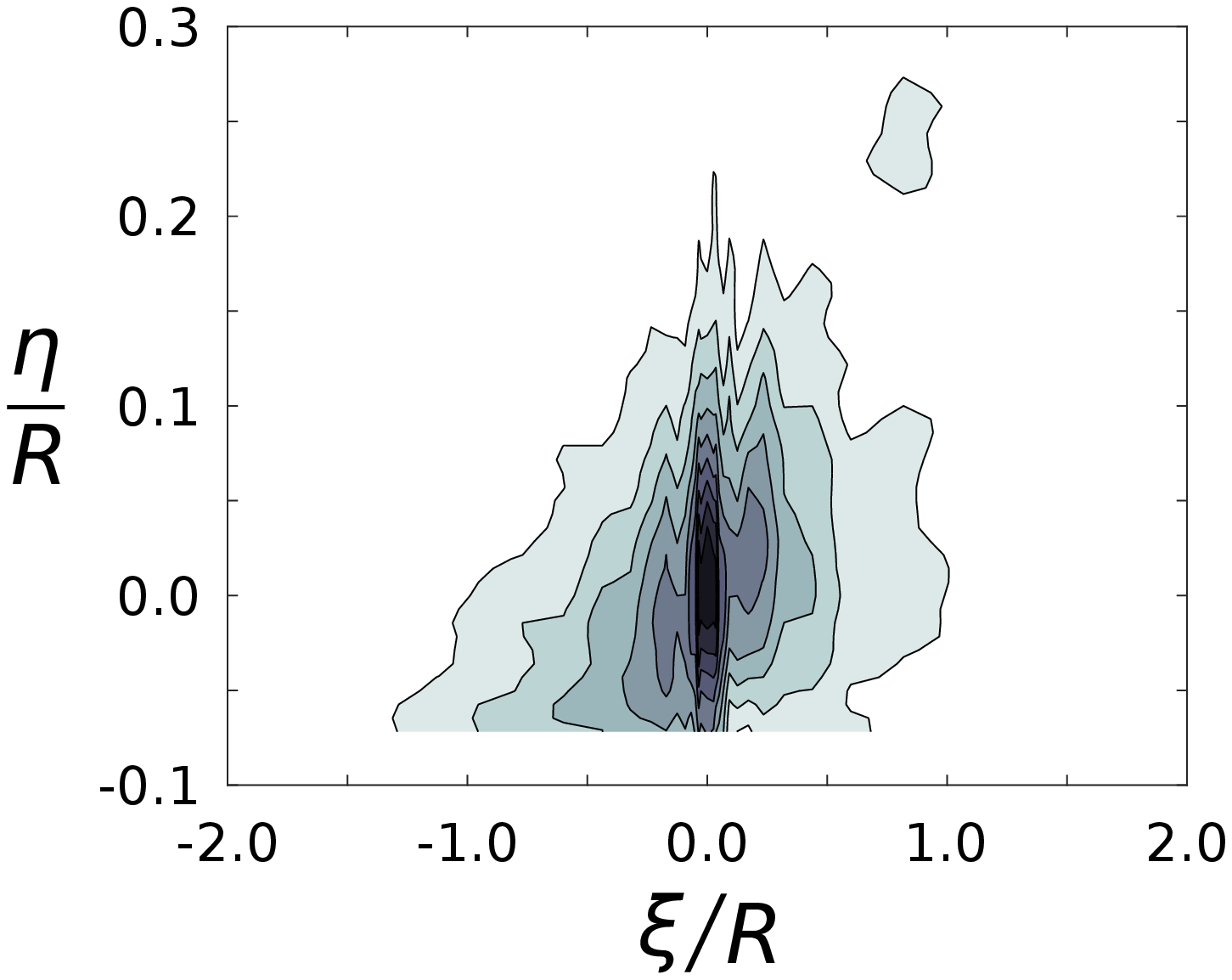}} &
			\sidesubfloat[]{\includegraphics[width=0.28\textwidth]{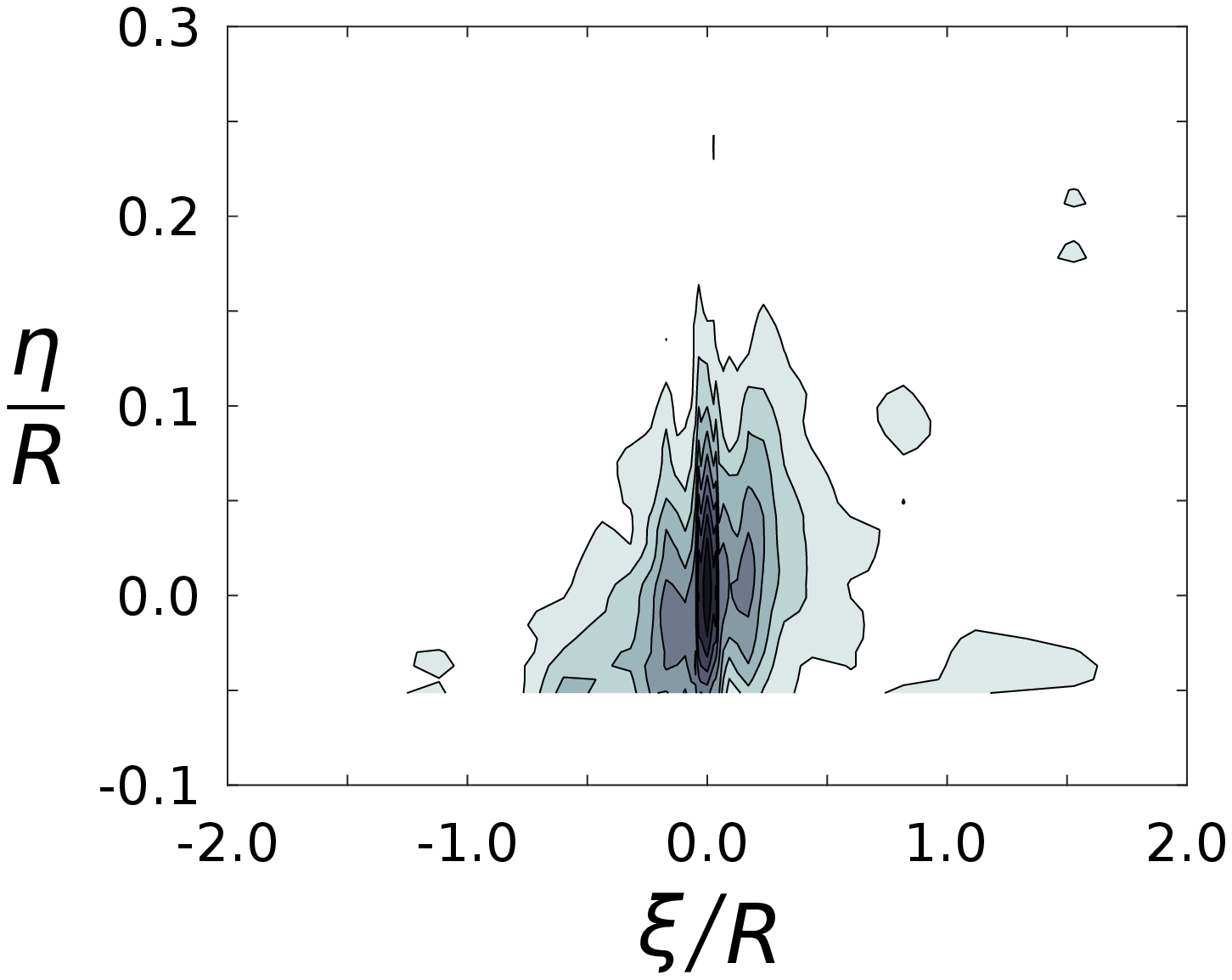}} \\
			\sidesubfloat[]{\includegraphics[width=0.28\textwidth]{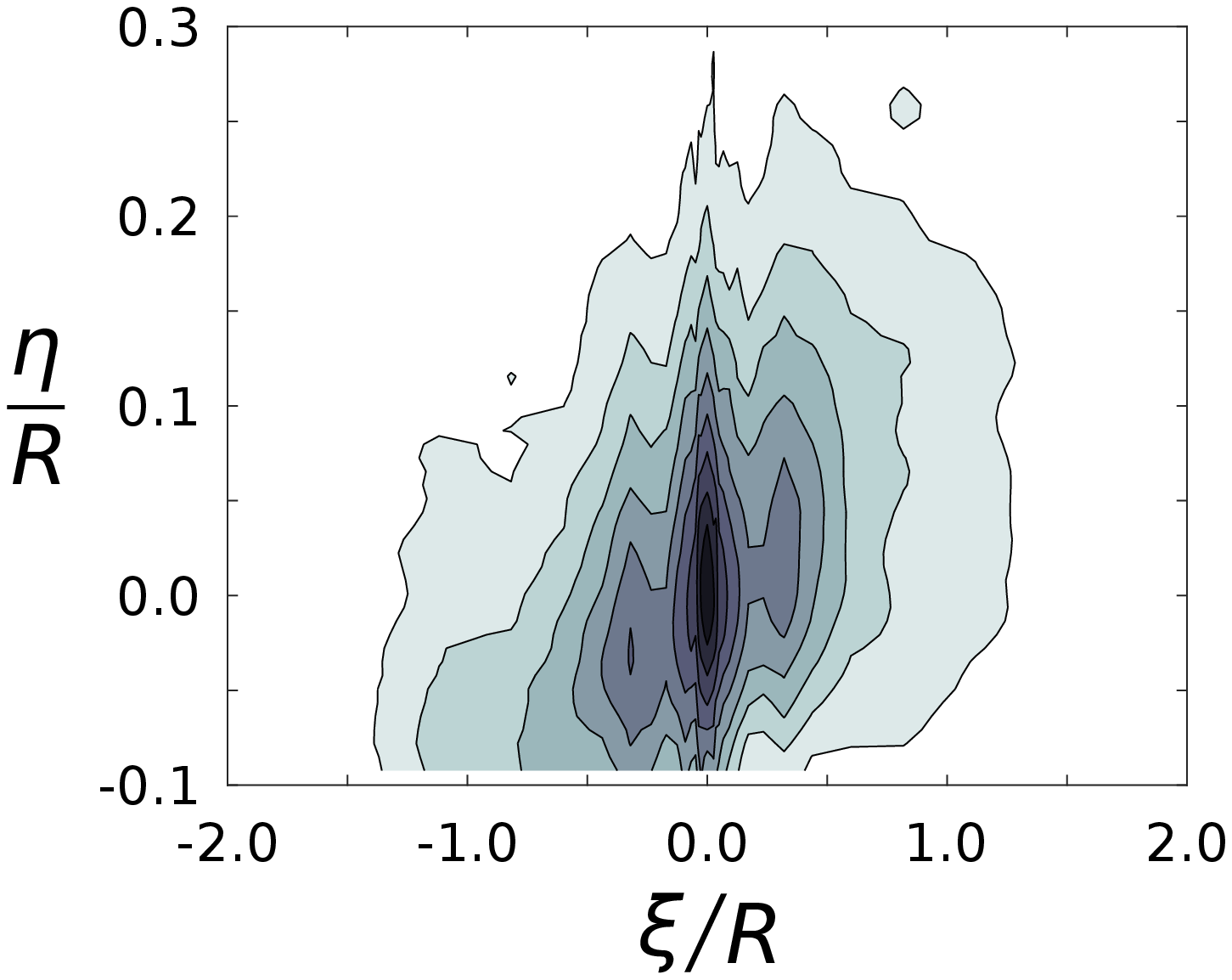}} &
			\sidesubfloat[]{\includegraphics[width=0.28\textwidth]{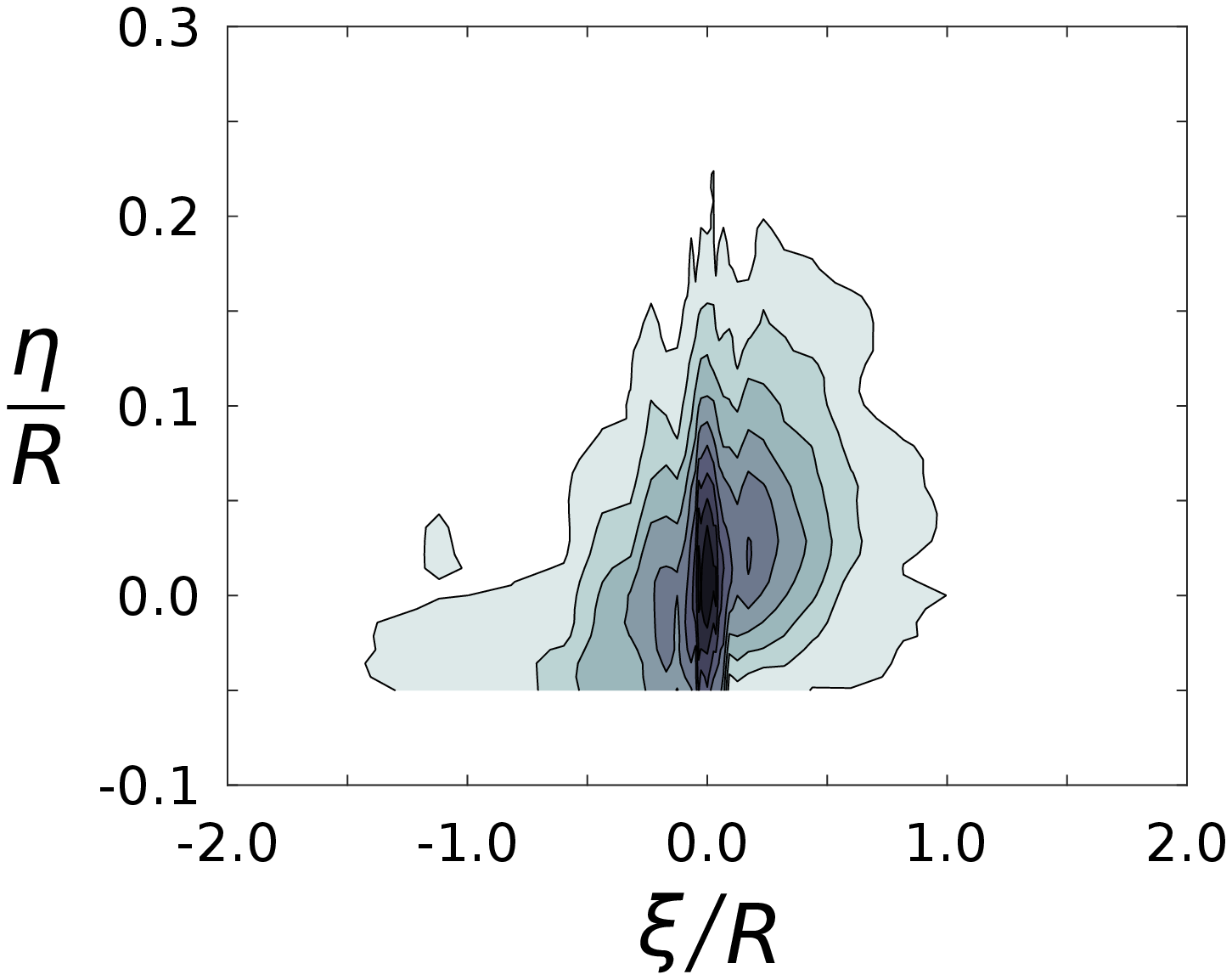}} &
			\sidesubfloat[]{\includegraphics[width=0.28\textwidth]{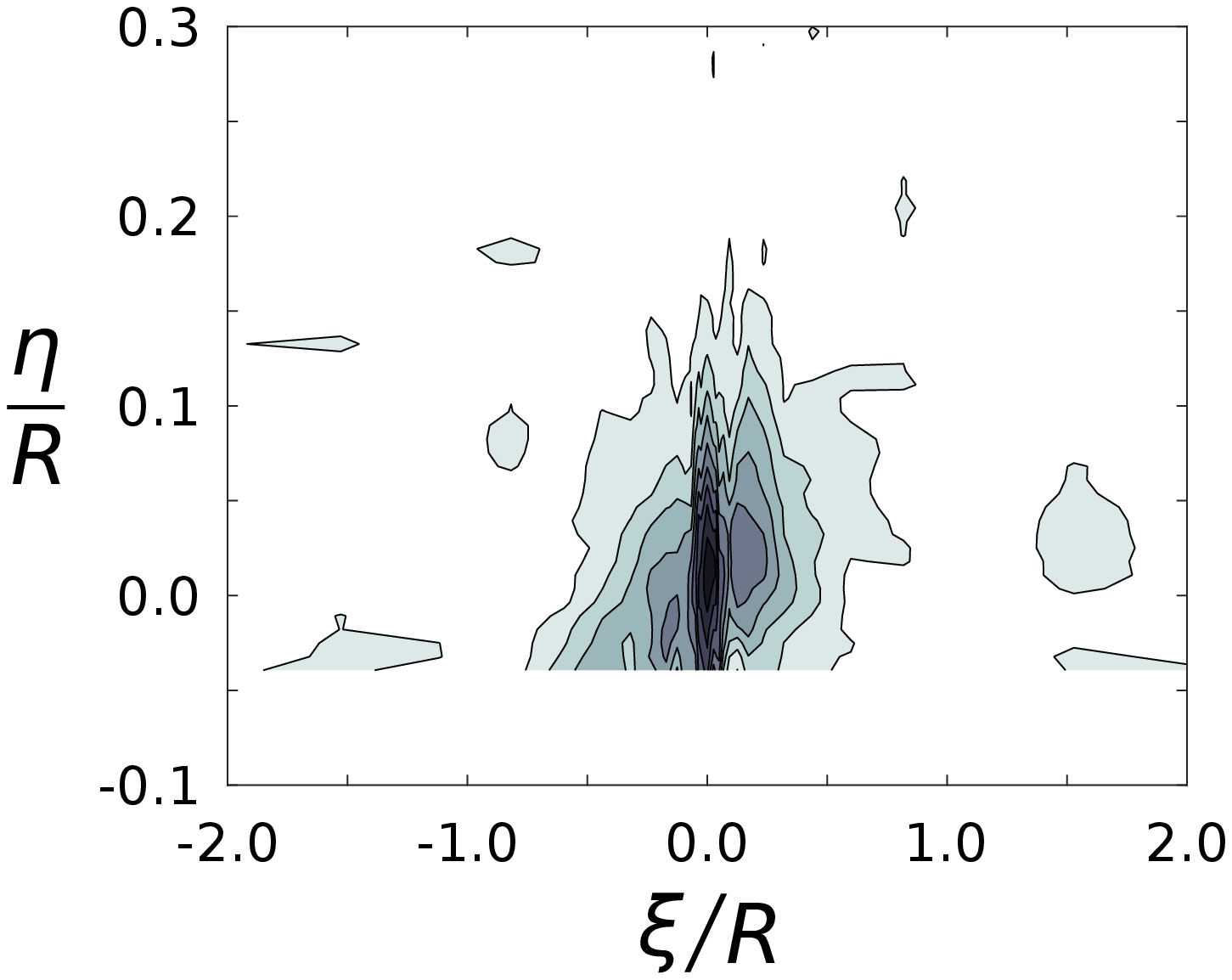}}
		\end{tabular}
 	\end{center}
	\caption{Absolute value of the correlation maps described in equation \ref{eq:corr}. The wall-normal reference point $y_{MAX}$ is located at the origin, with the wall approaching the origin for decreasing structure sizes. (a) $Re_\tau = 1310$, $m = 10$. (b) $Re_\tau = 1310$, $m = 15$. (c) $Re_\tau = 1310$, $m = 20$. (d) $Re_\tau = 2430$, $m = 10$. (e) $Re_\tau = 2430$, $m = 15$. (f) $Re_\tau = 2430$, $m = 20$. (g) $Re_\tau = 3810$, $m = 10$. (h) $Re_\tau = 3810$, $m = 15$. (i) $Re_\tau = 3810$, $m = 20$. Each correlation map is scaled by its peak value and are shown with contour lines ranging from 0.1 to 0.9 with 0.1 increments.}
\label{fig:corrMap}
\end{figure}

The absolute values of the correlation for  $m\in\{10,~15,~20\}$ are mapped in figure \ref{fig:corrMap}. All correlations were constructed using points from the two independent planes, including the auto-correlations ($\xi=0$), allowing any sampling noise to average out. The reference planes were inter-exchanged, such that we could construct both the upstream and downstream sections of the correlation maps. The correlation maps reveal a streamwise leaning structure, similar to the large-scale motions, with a decrease in height and length as their width decreases.

\begin{figure}
 	\begin{center}
		\begin{tabular}{ccc}
			\sidesubfloat[]{\includegraphics[width=0.28\textwidth]{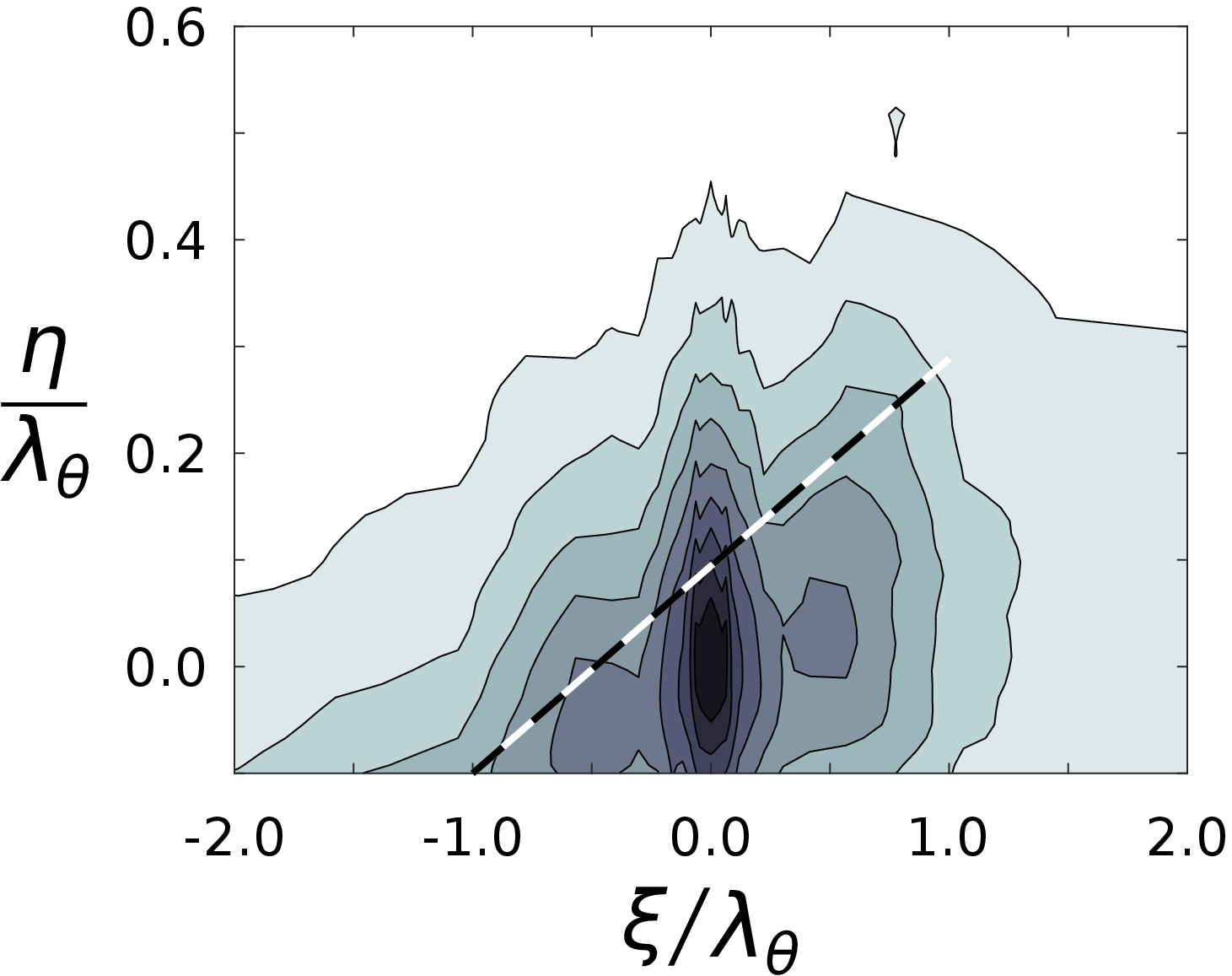}} &
			\sidesubfloat[]{\includegraphics[width=0.28\textwidth]{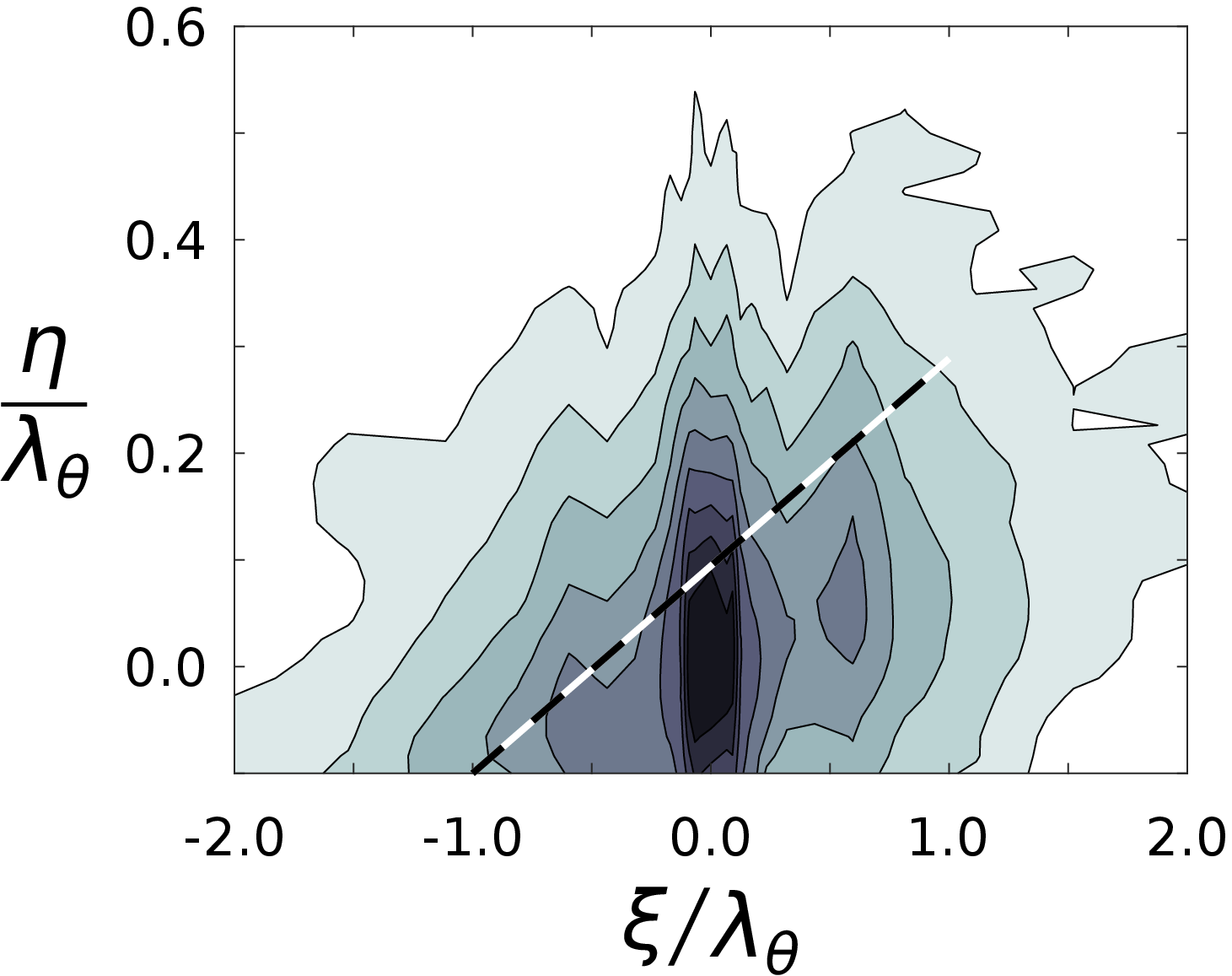}} &
			\sidesubfloat[]{\includegraphics[width=0.28\textwidth]{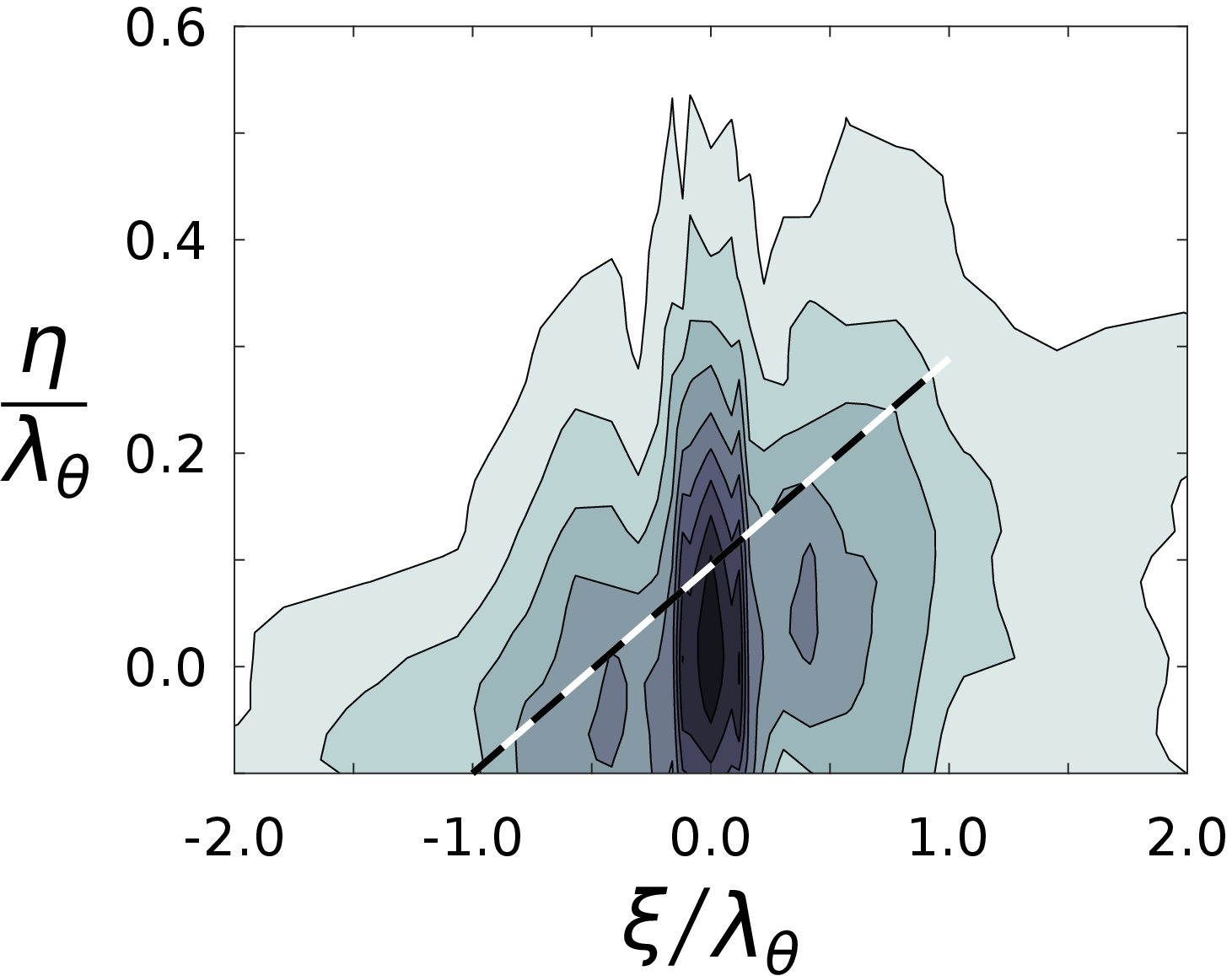}} \\
			\sidesubfloat[]{\includegraphics[width=0.28\textwidth]{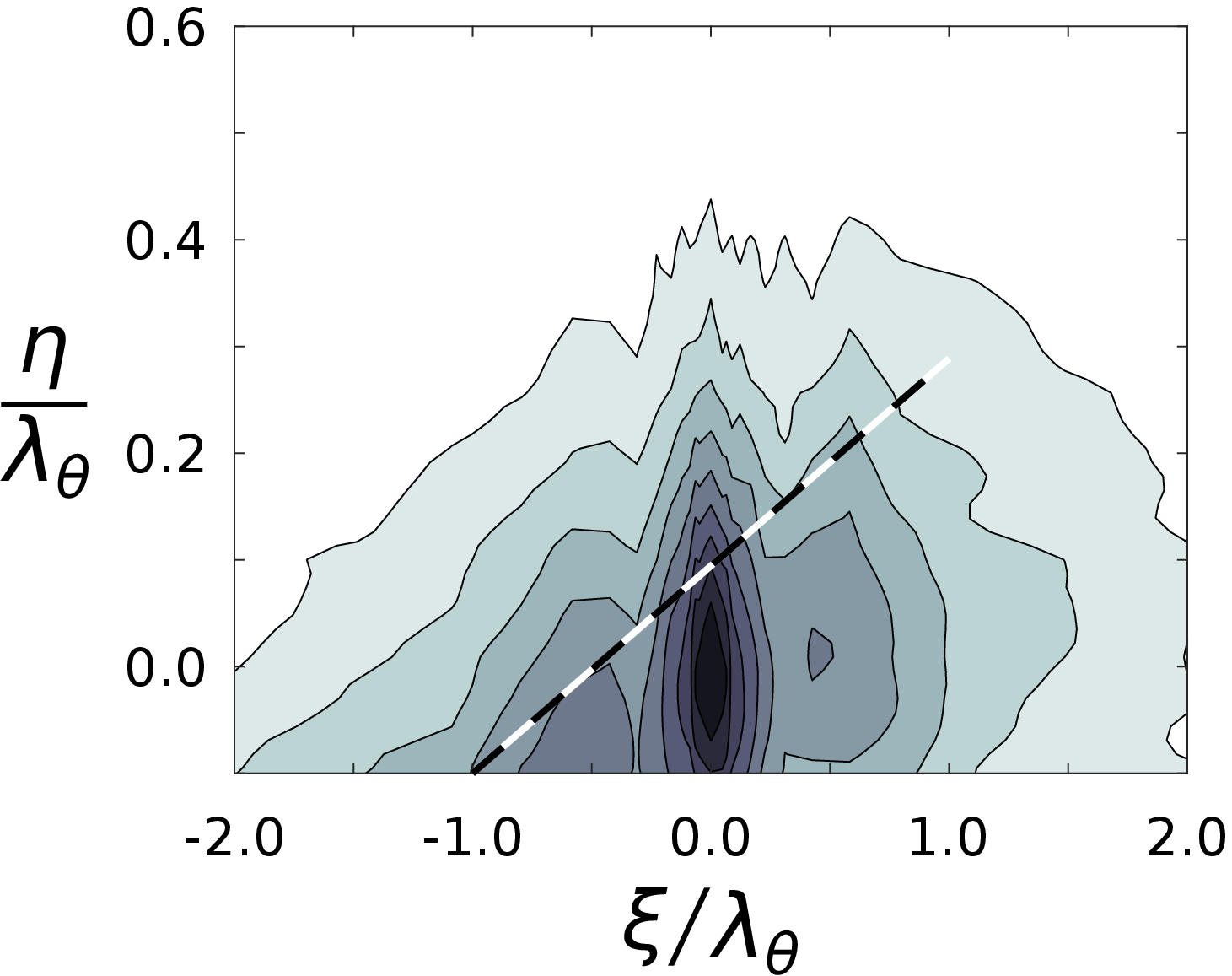}} &
			\sidesubfloat[]{\includegraphics[width=0.28\textwidth]{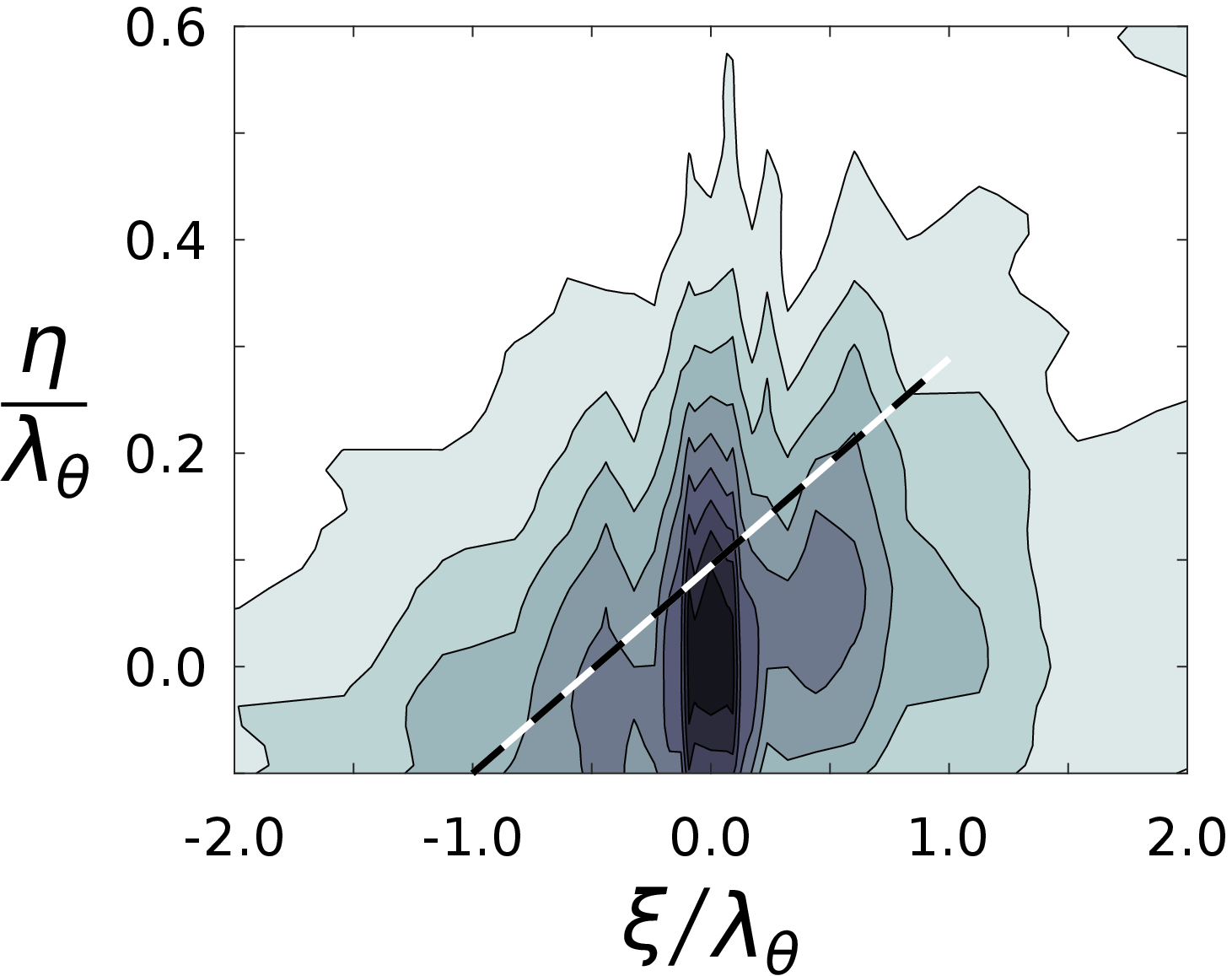}} &
			\sidesubfloat[]{\includegraphics[width=0.28\textwidth]{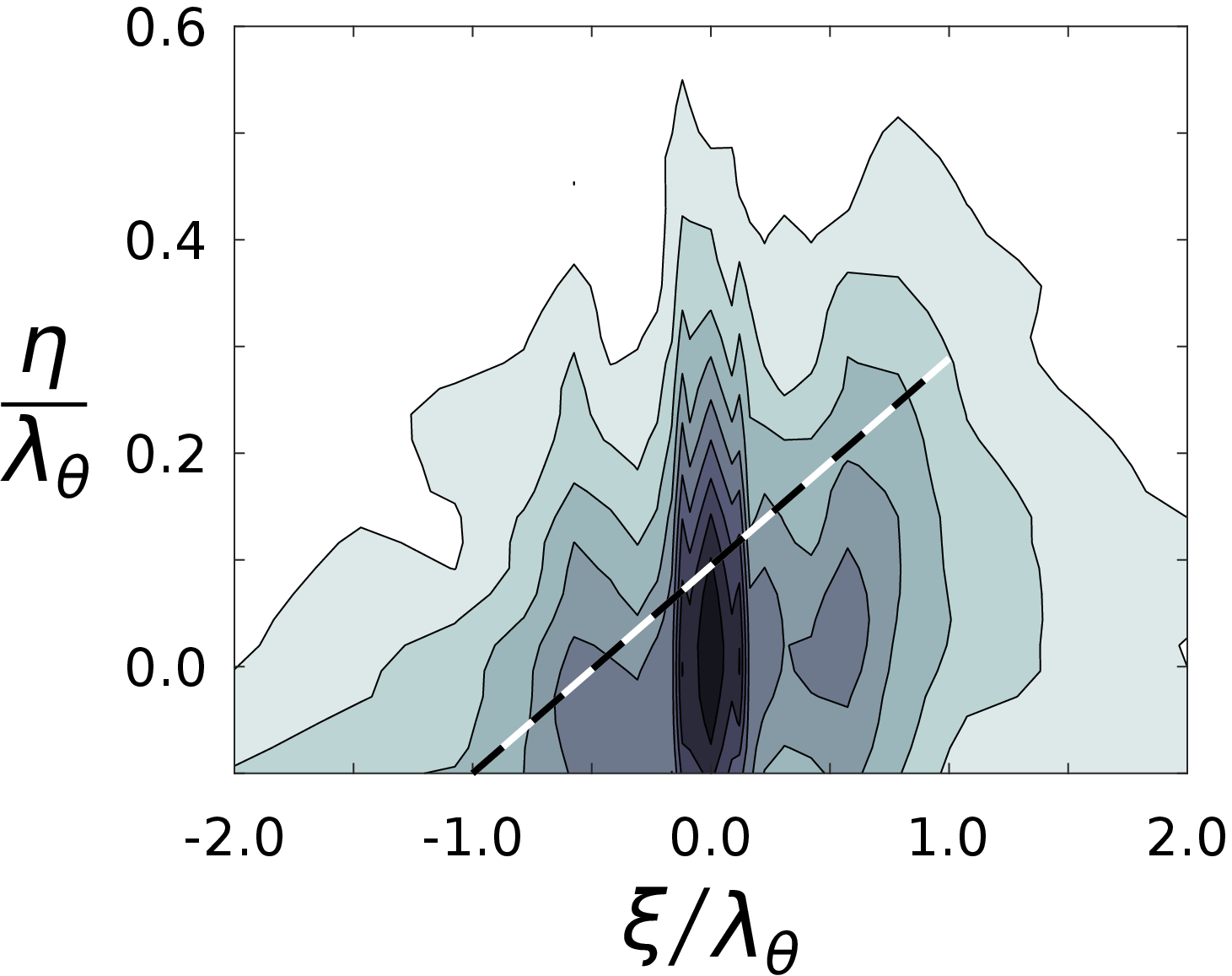}} \\
			\sidesubfloat[]{\includegraphics[width=0.28\textwidth]{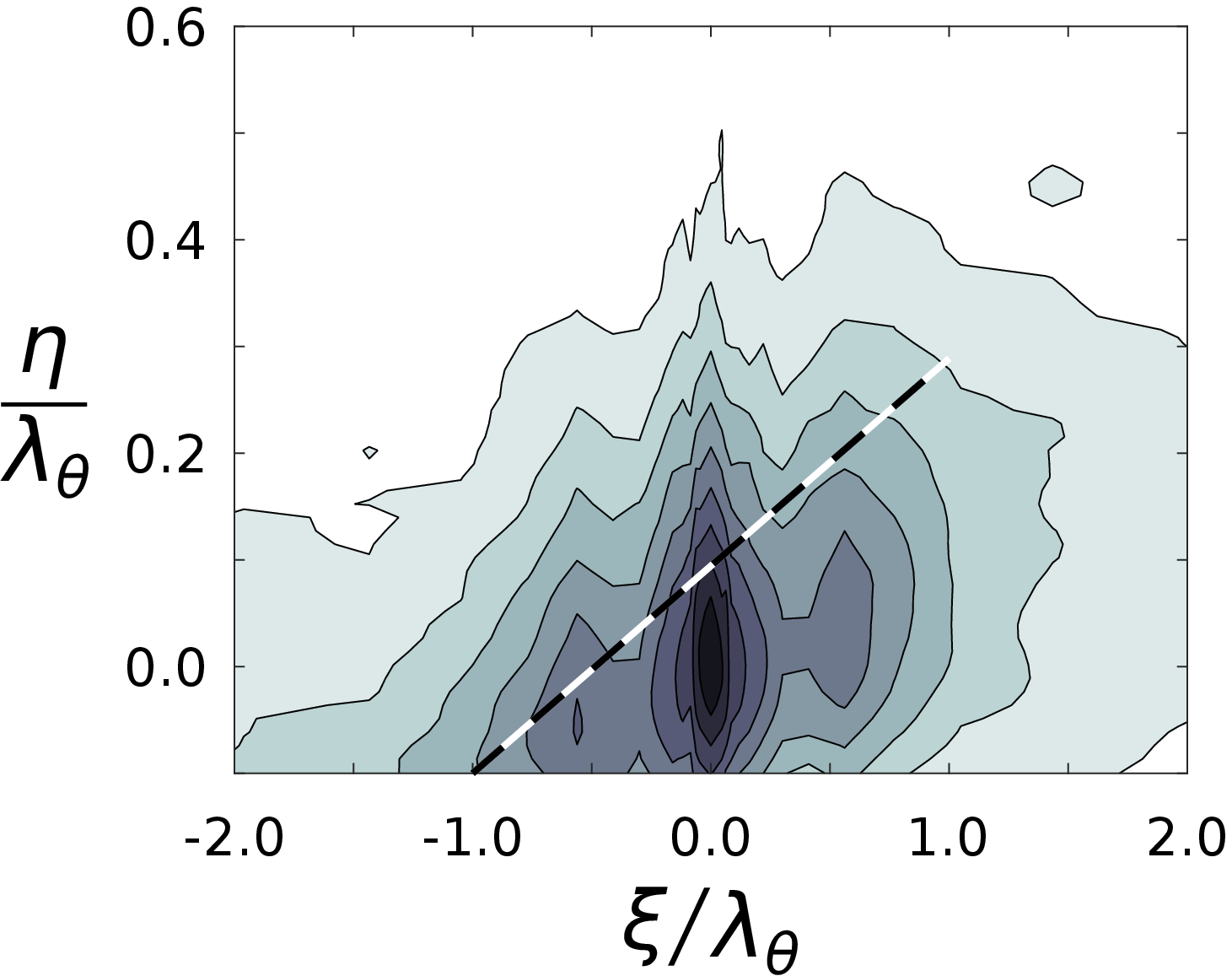}} &
			\sidesubfloat[]{\includegraphics[width=0.28\textwidth]{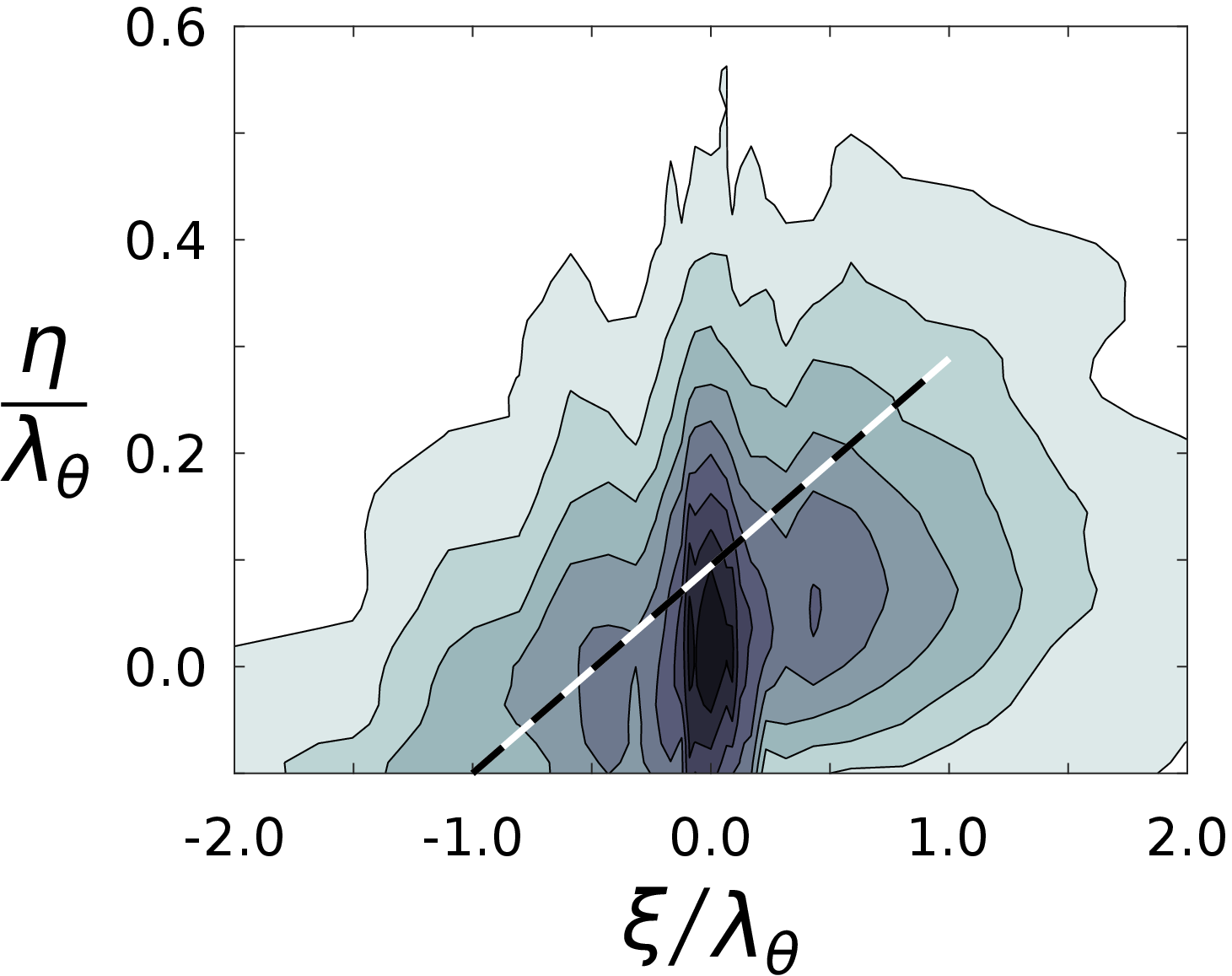}} &
			\sidesubfloat[]{\includegraphics[width=0.28\textwidth]{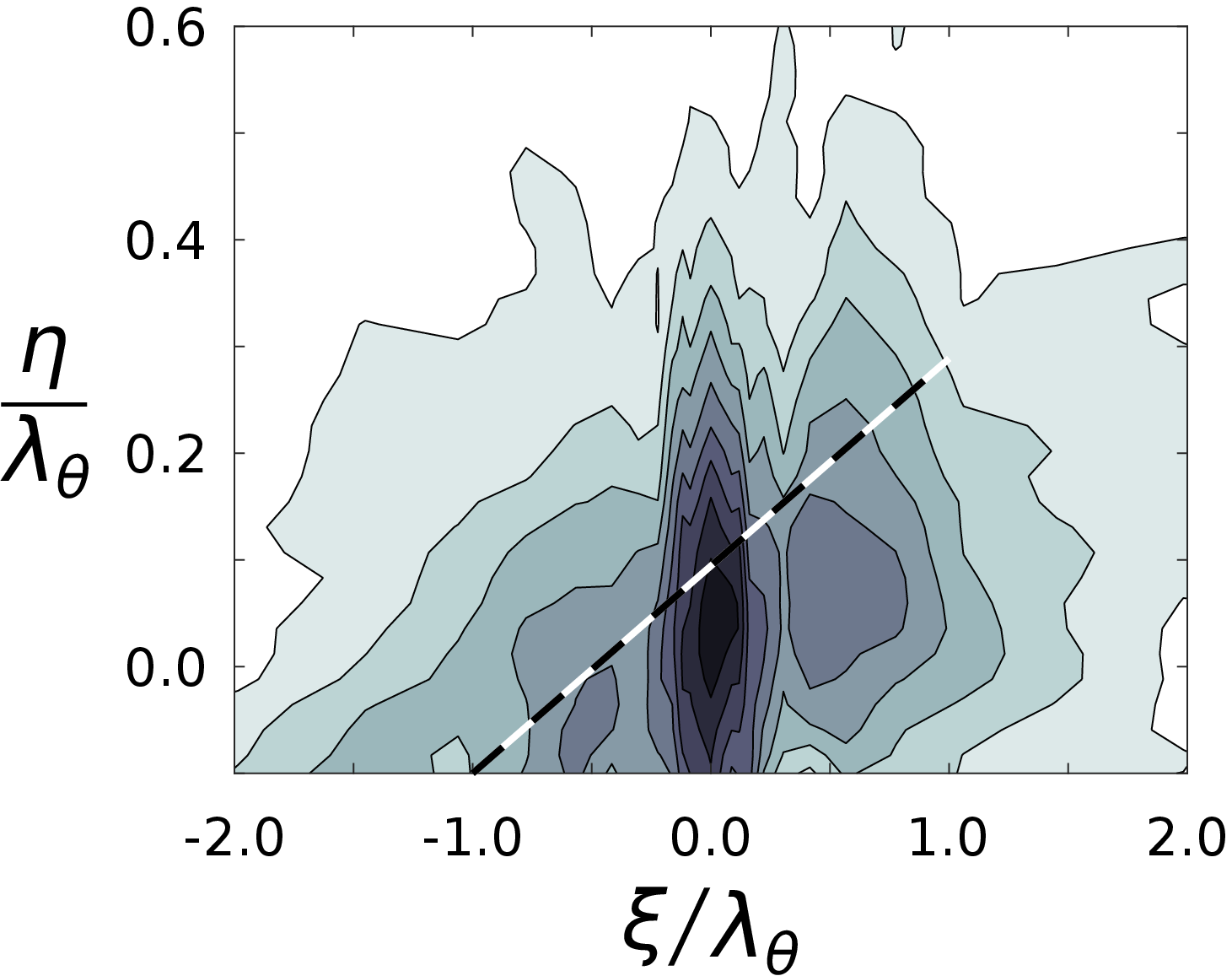}}
		\end{tabular}
 	\end{center}
	\caption{Absolute value of the scaled correlation maps. The wall-normal reference point $y_{MAX}$ is located at the origin, with the wall approaching the origin for decreasing structure sizes. (a) $Re_\tau = 1310$, $m = 10$. (b) $Re_\tau = 1310$, $m = 15$. (c) $Re_\tau = 1310$, $m = 20$. (d) $Re_\tau = 2430$, $m = 10$. (e) $Re_\tau = 2430$, $m = 15$. (f) $Re_\tau = 2430$, $m = 20$. (g) $Re_\tau = 3810$, $m = 10$. (h) $Re_\tau = 3810$, $m = 15$. (i) $Re_\tau = 3810$, $m = 20$. Each correlation map is scaled by its peak value and are shown with contour lines ranging from 0.1 to 0.9 with 0.1 increments. Dashed line represents an $11^\circ$ slope, as previously seen for LSMs \citep{Adrian2000}.}
\label{fig:corrMap_scaled}
\end{figure}

Figure \ref{fig:corrMap_scaled} shows the same correlation maps given in figure \ref{fig:corrMap}, but scaled using the eddy length scale $\lambda_\theta$. In these coordinates, the structures exhibit a convincingly self-similar behavior, where the structure height and length are approximately equal. 
We can also see a weak repetition within each structure, represented by a secondary (weaker) correlation peak located at $\xi/\lambda_\theta=\pm 0.5$.

The observed secondary peaks are consistent with the proposal that the large-scale motions (LSMs) are constructed from a set of streamwise aligned hairpin vortices \citep{Adrian2000}. 
It can also be seen that, in accordance with \citet{Adrian2000}, the upstream correlation peak at $\xi/\lambda_\theta= - 0.5$ is attached to the wall, while the downstream correlation peak at $\xi/\lambda_\theta=+ 0.5$ has detached and could be considered a more mature structure.
\cite{Adrian2000} further found that the typical slope of the ceiling of the LSM was $11^\circ$, subject to some variation due to size and maturity of the LSM, with larger and more mature structures exhibiting a steeper slope. Figure \ref{fig:corrMap_scaled} shows that the inclination for each identified structure is close to  11$^\circ$, again supporting the theory that the identified structures are the LSMs.

\begin{figure}
	\begin{center}
	\includegraphics[width=0.5\textwidth]{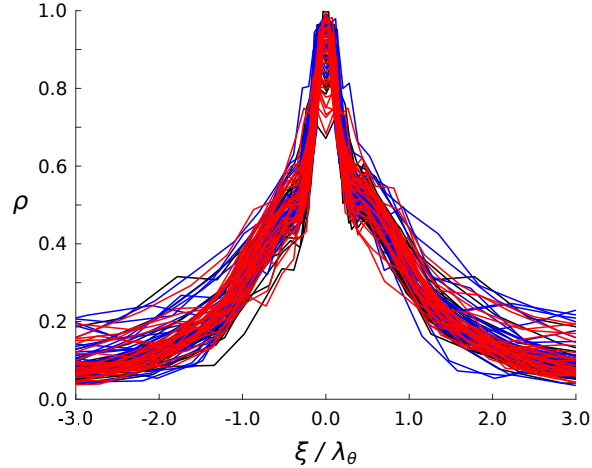}
	\end{center}
\caption{Absolute value of the correlation maps, evaluated at the eddy-specific wall-distance $y_{MAX}$, where the eddy velocity profile has its maximum.
Showing all profiles for the range $m \in [5,~35]$. $Re_\tau = \{1310,~2430,~3810\}$ are shown in black, blue and red, respectively. }
\label{fig:xCorr}
\end{figure}

Rather than relying on visual comparisons based on figure~\ref{fig:corrMap_scaled}, a more objective measure of the self-similarity in the streamwise direction is to compare the streamwise correlation of each eddy, evaluated at the eddy-specific wall distance $y_{MAX}$. Figure \ref{fig:xCorr} shows the correlations for all azimuthal modes in the range $m \in [5,~35]$. We have refrained from enforcing symmetry on the correlation profiles, and their symmetry can be seen as indicating a well converged data set. There is a clear similarity between the correlation profiles, although there is some spread seen in the tails.  There is no distinguishable difference over the range of Reynolds numbers considered here. 

\begin{figure}
	\begin{center}
	\includegraphics[width=0.7\textwidth]{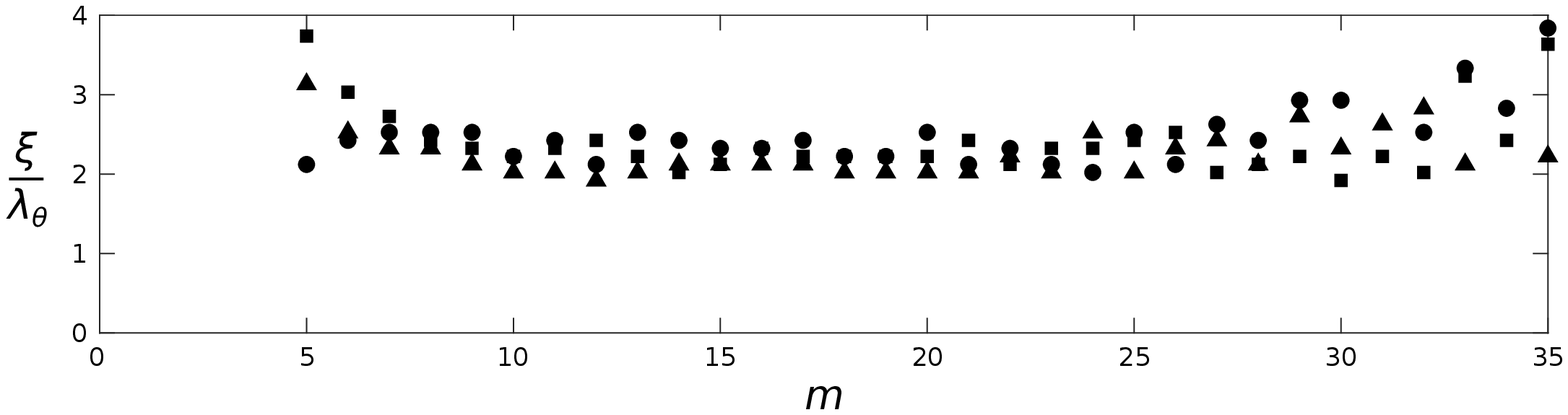} \\
	\includegraphics[width=0.7\textwidth]{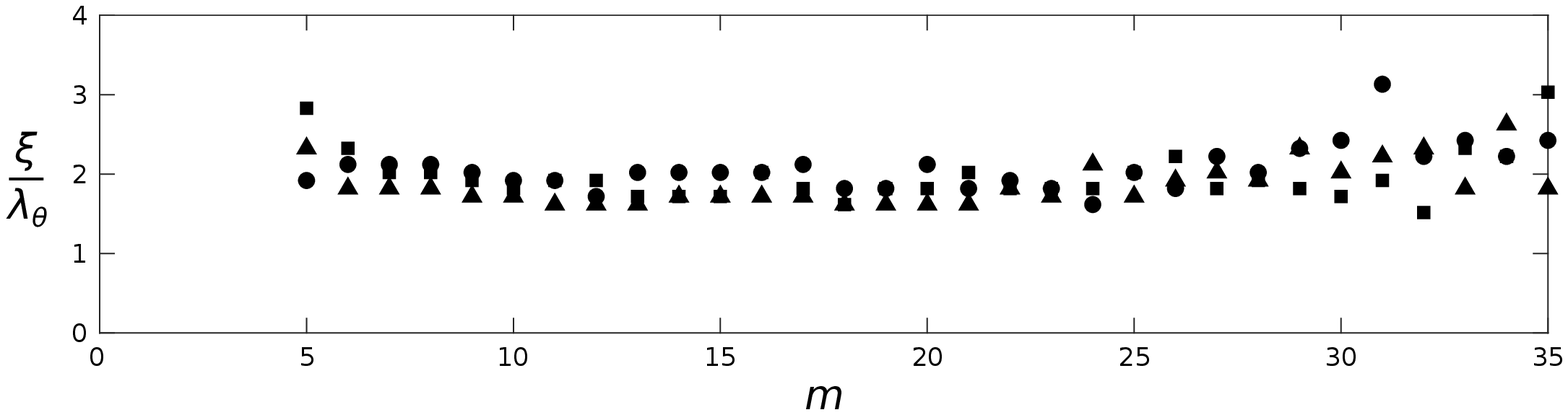} \\
	\includegraphics[width=0.7\textwidth]{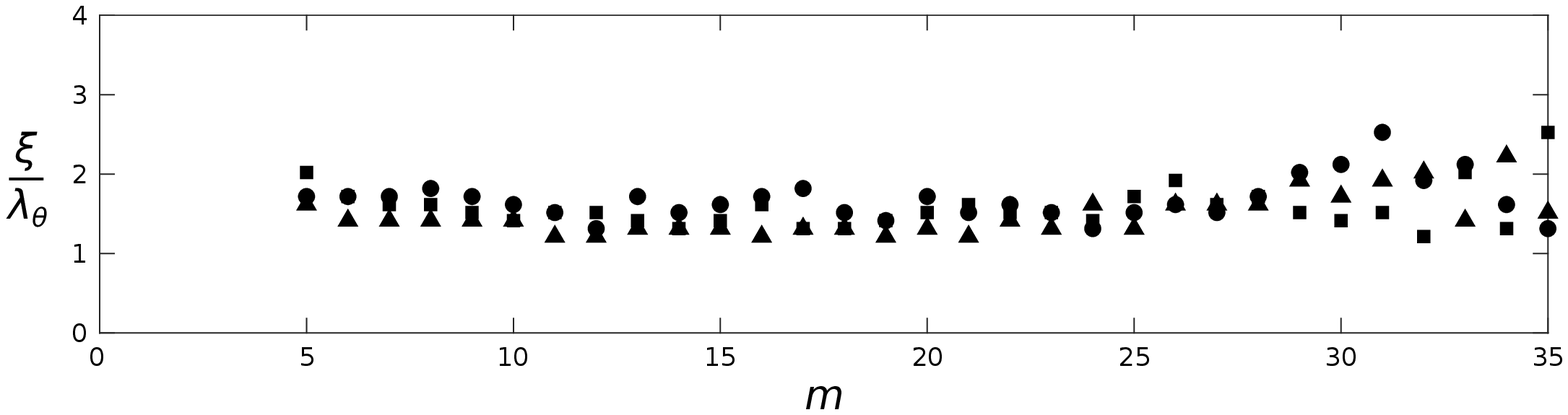} 
	\end{center}
\caption{Streamwise correlation lengths, evaluated from the profiles in figure \ref{fig:xCorr}. The considered thresholds are: (a) $0.08$, (b) $0.10$, (c) $0.12$. $\blacktriangle$ $Re_\tau = 1310$; $\blacksquare$ $Re_\tau = 2430$; $\bullet$ $Re_\tau = 3810$.}
\label{fig:xCorr_thold}
\end{figure}

In order to examine the correlation tails in more detail, we now determine the streamwise eddy length as the distance between the upstream and downstream ends of the correlation where the correlation is above a chosen threshold.  Figure~\ref{fig:xCorr_thold} shows the streamwise length of each eddy, normalized by its width, for three different  thresholds, $0.08$, $0.10$ and $0.12$. The scaled eddy length remains reasonably constant for $m \in[6,~30]$, and it is insensitive to eddy size, and the chosen threshold. This range of $m$ corresponds to eddy sizes ($\lambda_\theta/R$) between $[0.885,~0.203]$, $[0.870,~0.203]$ and $[0.890,~0.203]$ for Reynolds numbers 1310, 2430 and 3810, respectively. Due to the nature of the Fourier transforms, these lengths corresponds to a positive structure and its negative neighbor, and each structure width should be taken as $\lambda_\theta /2$, if we take the width of a LSM to be the distance between its widest legs.   However, the departure from this scaling for the smaller eddies is most likely due to the limited streamwise resolution along with the volumetric velocity averaging in the PIV process. For instance, the smallest eddy width of 0.2$R$ in the current setup corresponds to $3.8$~mm, compared to the $1$~mm thick laser sheet. It is therefore possible that the self-similar behavior might extend to even smaller structures if investigated using different data acquisition tools.

%------------------------------------------------------
% Conclusion ------------------------------------------
%------------------------------------------------------
\section{Discussion and Conclusions}

The eddy scaling of fully-developed turbulent pipe flow was investigated in the context of Townsend's attached eddy hypothesis.  The eddy structure was identified using Fourier decomposition in the spanwise plane, and two-point correlations in the streamwise plane. In the spanwise plane, the average eddy velocity profile corresponding to a given azimuthal mode was created by azimuthally aligning the structures. The eddy velocity profiles displayed an inner and outer region, where the outer region obeyed the classical outer layer scaling using pipe radius and the friction velocity. For the inner region, the height of the eddy velocity profiles, for $m \in [5,~35]$, were scaled by their characteristic width, as expected for attached eddies. Within this range, the structure sizes ranged approximately $\lambda_\theta/R \in [1.03,~0.175]$, where the length scale ratio was $5.89$. The eddy velocity magnitude, however, decreased with eddy size, approximately in proportion to both friction velocity and eddy size. In the streamwise direction, the two-point correlation showed that the streamwise length of these eddies scaled using their characteristic width so that the structures display a fully self-similar behavior (for $m \in[6,~30]$. The self-similar range can also be expressed as $\lambda_\theta/R \in [0.885,~0.203]$, $[0.870,~0.203]$ and $[0.890,~0.203]$, for Reynolds numbers $1310$, $2430$ and 3810, respectively. For this structure range, the resolved length scale ratio is $4.3$, while the volumetric ratio is $1:80$. The correlations also revealed the presence of smaller repetitive structures within the larger comportment, in accordance with the expected behavior of the large scale motions. The detected structures, showed no significant Reynolds number effects over the range of Reynolds numbers studied here.

We would, finally, like to note that Townsend proposed a simplified framework for how turbulence can be viewed. It is a linear model and all non-linearities must either be neglected or included in the base structure. Because this is one of the fundamental assumptions of the attached eddy model, one must posit that a potent base-structure should include the non-linearities, and consequently must also be allowed to be composed of a range of smaller structures, as suggested by the findings in this work.

This work was supported under ONR Grants N00014-15-1-2402 (Program Manager Ron Joslin) and N00014-17-1-2309 (Program Manager Joseph Gorski).

%------------------------------------------------------
% References ------------------------------------------
%------------------------------------------------------
\bibliographystyle{refclass}

\end{document}